\documentclass{article}
\usepackage{amsmath, amsthm}
\usepackage{graphicx,epsfig,epic}
\font\tenbb=msbm10 at 12pt

\def\rR{\hbox{\tenbb R}}

\def\cal{\mathcal}

\def\gesp{\vskip1cm}
\def\esp{\vskip .6cm}
\def\pesp{\vskip .3cm}

\def\ni{\noindent}
\def\di{\displaystyle}

\newtheoremstyle{theorem}
{10pt} 
{10pt} 
{\sl} 
{\parindent} 
{\bf} 
{. } 
{ } 
{} 
\theoremstyle{theorem}
\newtheorem{theorem}{Theorem}
\newtheorem{corollary}[theorem]{Corollary}
\newtheorem{lemma}[theorem]{Lemma}
\newtheorem{proposition}[theorem]{Proposition}
\newtheoremstyle{defi}
{10pt} 
{10pt} 
{\rm} 
{\parindent} 
{\bf} 
{. } 
{ } 
{} 
\theoremstyle{defi}
\newtheorem{definition}[theorem]{Definition}

\begin{document}
\title{Mathematical Model for Fractal Manifold}
\author{Fay\c cal BEN ADDA\\
University of Hail, Department of mathematics,\\
 P. O. Box 2440, Hail, KSA\\E-mail: fbenadda@uoh.edu.sa}
\maketitle
\begin{abstract}
We have built a new kind of manifolds which leads to an
alternative new geometrical space. The study of the nowhere
differentiable functions via a family of mean functions leads to a
new characterization of this category of functions. A fluctuant
manifold has been built with an appearance of a new structure on
it at every scale, and we embedded into it an internal structure
to transform its fluctuant geometry to a new fixed geometry. This
approach leads us to what we call fractal manifold. The elements
of this kind of manifold appear locally as tiny double strings,
with an appearance of new structure at every step of
approximation. We have obtained a variable dimensional space which
is locally neither a continuum nor a discrete, but a mixture of
both. Space acquires a variable geometry, it becomes explicitly
dependent on the approximation process, and the geometry on it
assumed to be characterized not only by curvature, but also by the
appearance of new structure at every step of approximation. {\bf
AMS Subject Classification:} 58A03, 58A05, 28A80. \\{\bf Key Words
and Phrases:} Topos-theoretic approach to differentiable
manifolds, Differential geometry foundations, Fractals.
\end{abstract}
\section{Introduction}
Up to now  there is no real understanding of the nature of the
space of an elementary particle, and the current space-time is not
satisfactory for particle theory \cite{BA}. Moreover, Feynman has
demonstrated that the typical quantum mechanical paths (that
contribute in a dominant way to the path integral) are continuous
but nowhere differentiable, and can be characterized by a fractal
dimension 2 \cite{FE1,FE2,SC}. A fractal\footnote{ The word {\it
fractal} is a new term introduced by  Mandelbrot \cite{MA1,MA2} to
represent shapes or phenomena having no characteristic length.}
space time was suggested in 1983 by G. Ord and L. Nottale
\cite{N4,OR}, but until now there is no mathematical model of such
space. Given up the differentiability\footnote{ In a letter to
Pauli \cite{EI}, Einstein suggested that a true understanding of
quantum physics could imply to give up differentiability.} to
describe the space-time in quantum mechanics was suggested early
by Einstein without pursuit, and suggested in 1993 by Nottale
\cite{N2}, which leads to the scale relativity. Giving up the
differentiability means that we have to deal with the nowhere
differentiable functions, and the difficulties encountered are as
follow:

i) There are no much tools nor analytic characterization for the
nowhere differentiable functions \cite{BC,TR} other than the
notion of fractal dimension.

ii) There is no definition of nowhere differentiable manifolds to
deal with it, and this is a big challenge. Do we need a class of
manifolds with no smooth structure on it? Or a class of manifolds
with a smooth structure on it? Or any other kind of mathematical
object which gives us a variable geometrical space?

A non differentiable manifold with no smooth structure on it is an
old problem at the time of Milnor, unsolved because of its
difficulty. Our purpose in the present paper is to investigate the
possibility to construct a manifold that presents a variable local
geometry. For this purpose our plan is as follow:

a) We approximate the nowhere differentiable function by a mean
function which follows the geometry of the approximated function

b) We introduce a new characterization of nowhere differentiable
functions, and we define a double space called
($\varepsilon$-manifolds) using graphs.

c) We introduce a model of fractal manifold using family of
graphs.

e) At the end, using the process mean of the mean we prove that
elements of the fractal manifold are double strings with
appearance of new structures on it at every step of approximation.
The fractal manifolds obtained are locally neither a continuum nor
a discrete, but a mixture of both.

\section{Basic Concepts and Tools }
\subsection{Scale and Resolution}
Let $[a,b]$, $a<b$,  be an interval included in $\rR$, with length
$l=b-a$. It is possible to construct a family of subintervals of
length $l= {b-a\over n}$, $\di\forall n\geq 1$. This kind of
subdivision is countable and the subintervals obtained depend on
the real numbers $a$ and $b$. For a non countable subdivision of a
unit interval, let us consider the length $l=\di{1\over \beta}$,
$\beta>0$ real number and we introduce the following definitions.
\begin{definition}\label{R}
We call {\it resolution} the quotient of the unit by a real number
$\beta>0$ and we denote it by  $ \varepsilon=\di{1\over
\beta},\quad\hbox{for all}\quad\beta>0.$
\end{definition}
This resolution $\varepsilon$ takes an infinity of values between
0 and $+\infty$, and we can extend it by continuity on zero. We
admit that the resolution $\varepsilon$ is equal to zero as
$\beta$ tends to infinity and we have $\forall \beta>0$,
$\varepsilon\in[0,+\infty[$.

\begin{definition}
We call {\it scale} and we denote it by $\cal E$ the inverse of
the resolution, such that ${\cal E}\varepsilon=1$.
\end{definition}

Let us consider a continuous and nowhere differentiable function
$f(x)$,  in a given interval ${\cal I}\subset\rR $. Since $\forall
x\in {\cal I}$ we can not find ${df(x)\over dx}$, it is impossible
to draw the graph of this kind of functions. However, it's
possible to approximate it through differentiable functions. We
replace the nowhere differentiable function $f$ by the following
mean function given by
\begin{equation}
\label{M}
 f(x,\varepsilon)=\di{1\over
2\varepsilon}\int^{x+\varepsilon}_{x-\varepsilon}f(t)dt,\qquad\forall
x\in {\cal I}.
\end{equation}

\subsection{Approximation Domain}

 For all $\varepsilon>0$, the derivative of the mean function
(\ref{M}) is given by
$f'(x,\varepsilon)={f(x+\varepsilon)-f(x-\varepsilon)\over
2\varepsilon}$. For $\varepsilon=0$ we have
$f(x,0)=\lim_{\varepsilon\to 0}f(x,\varepsilon)=f(x)$ which is
nowhere differentiable, and $f(x,\varepsilon)$ is always
differentiable for all $\varepsilon\in]0,+\infty[$. We call domain
of approximation of the function $f$ the set
\begin{equation}
{\cal Q }_f=\{\ \varepsilon\in\rR{^+}\ /\
f(x,\varepsilon)\quad\hbox{is differentiable on ${\cal I}$}\ \},
\end{equation}

\begin{definition}
We call small resolution domain, and we denote it by ${\cal R}_f$,
the intersection between the domain of approximation ${\cal Q }_f$
and $[0,\alpha]$, $0<\alpha\ll1$
\begin{equation}
{\cal R }_f=\{\ \varepsilon\in\rR{^+}\ /\ f(x,\varepsilon)
\quad\hbox{is differentiable on ${\cal I}$}\ \}\cap[0,\alpha].
\end{equation}
\end{definition}

\subsection{Rate of Change}

Let us  consider a function $f(x)$, $x\in\cal I\subset\rR $ an
open interval, where the only information we have is given by the
following:

1) The function $f$ is continuous on $\cal  I $.

2) The function $f$ is nowhere differentiable on $\cal  I $.

3) The function $f$ is unknown.

4) The mean function given by (\ref{M}) is defined for all
$\varepsilon\in {\cal Q }_f$.

What kind of information can we obtain about the unknown function
$f$? In this case, we can't say that the main information we have
is given by the best approximation, and this for the simple reason
that we can't determine the error of approximation if the function
is unknown. One can say by experience that the best approximation
is given by using the smallest resolution, but the problem here is
as follow: there is no smallest value of $\varepsilon\in]0,1]$ (no
minimal resolution in $]0,1]$). We propose the following lemma

\begin{lemma}\label{L1}
1) Let $f$ be a continuous and nowhere differentiable function. If
the mean forward $F^+(x,\varepsilon)=\di{1\over \varepsilon}\int
_x^{x+\varepsilon}f(t)dt$, and the mean-backward
$F^-(x,\varepsilon)=\di{1\over \varepsilon}\int
^x_{x-\varepsilon}f(t)dt$, $\forall(x,\varepsilon)\in I\times{\cal
R}_{f}$, then
\begin{equation}\label{E0}
f(x)=F^\sigma(x,\varepsilon)+\varepsilon\Big({\partial
F^\sigma(x,\varepsilon)\over\partial \varepsilon}-\sigma{\partial
F^\sigma(x,\varepsilon)\over\partial x}\Big),\qquad \sigma=\pm.
\end{equation}
2) Let $f$ be a continuous and nowhere differentiable function, if
\begin{equation}
F^\sigma(x,\delta_1,..,\delta_{n})={(\sigma1)^n\over\prod_{i=1}^n\delta_{i}}
\int_x^{x+\sigma\delta_{n}}
...\int_{t_1}^{t_1+\sigma\delta_1}f(t_0) d t_{0}..d t_{n-1},
\quad\sigma=\pm.
\end{equation}
$\forall\delta_2,..,\delta_{n}\in[0,\alpha]$, $0<\alpha\ll1$,
$\forall\delta_1\in{\cal R}_{f}$, $x\in{\cal I}$, then $\forall
n\geq1$, $\sigma=\pm$
\begin{equation}f(x)=F^\sigma(x,\delta_1,..,\delta_{n})+\sum_{i=1}^n\delta_i\Big ({\partial
F^\sigma(x,\delta_1,..,\delta_{i})\over\partial
\delta_i}-\sigma{\partial
F^\sigma(x,\delta_1,..,\delta_{i})\over\partial x}\Big ).
\end{equation}
\end{lemma}
\begin{proof}
1) It is not difficult to see that
\begin{equation}
\label{M3}
 {\partial F^+(x,\varepsilon)\over\partial
x}={f(x+\varepsilon)-f(x)\over\varepsilon},\quad {\partial
F^+(x,\varepsilon)\over\partial
\varepsilon}={-1\over\varepsilon}F^+(x,\varepsilon)+{f(x+\varepsilon)\over\varepsilon}.
\end{equation}

 From the equations (\ref{M3}), one can find
$$f(x)-F(x,\varepsilon)=\varepsilon\Big ({\partial
F(x,\varepsilon)\over\partial \varepsilon}-{\partial
F(x,\varepsilon)\over\partial x}\Big ).$$ Similar computation for
$\sigma=-$ allow us to conclude.

 2) Using
formula (\ref{E0}) and by induction over n we find the result.
\end{proof}
{\bf Remark}\label{R0} If we have the graphs of the mean forward
function or the mean backward function for all $(x,\varepsilon)\in
I\times{\cal R}_f$ of a nowhere differentiable function $f$, then
it is sufficient to study the variation of the mean forward
function or the mean backward function for all $(x,\varepsilon)\in
I\times{\cal R}_f$ to obtain the unknown function $f$. It means
that we must consider not only one approximation but all the
approximations given for all $\varepsilon\in{\cal R}_f$. We know
that the position of an electron and his momentum are subject to
Heisenberg's uncertainly principle. It states that the position
and momentum of a particle at the quantum microscopic level, can
not be measured simultaneously with complete accuracy. Hence, this
means in our case that it is impossible to determine our unknown
function by one approximation (even the best), actually, we can
not obtain a real description of the function $f$ if we consider
only one approximation\footnote{If we can draw the graph of one
mean function $F_1$, then we have a measurement of the position
$(x,F_1(x))$ and the momentum simultaneously (since we have
${dF_1\over dx}$).} out of all.

\subsection{Double and Multiple Additions}

 Let us consider the function given by the formula (\ref{M}) for
a given continuous and nowhere differentiable function $f(x)$,
$x\in{\cal I}$. This mean function can be written as sum of two
mean functions
\begin{equation}\label{d0}
f(x,\varepsilon)={1\over2}\Bigg\lbrack
f(x-{\varepsilon\over2},{\varepsilon\over2})+f(x+{\varepsilon\over2},
{\varepsilon\over2})\Bigg\rbrack
\end{equation}
where the functions $f(x-{\varepsilon\over2},{\varepsilon\over2})$
and $f(x+{\varepsilon\over2},{\varepsilon\over2})$ represent an
approximation of the function $f$ at the resolution
${\varepsilon\over2}$. The graph of the function
$f(x-{\varepsilon\over2},{\varepsilon\over2})$ denoted by
$\Gamma^-_\varepsilon$ is a copy of the graph of the function
$f(x+{\varepsilon\over2},{\varepsilon\over2})$ denoted by
$\Gamma^+_\varepsilon$, where $\Gamma^+_\varepsilon$ is obtained
by horizontal translation from the graph $\Gamma^-_\varepsilon$.
If we superimpose the two graphs $\Gamma^-_\varepsilon$ and
$\Gamma^+_\varepsilon$, we obtain two identical graphs of the
function $f$ with gap. This procedure can be repeated n times and
we have the following lemma

\begin{lemma}
Let $f$ be a continuous function on $\cal I$. For all
$(x,\varepsilon)\in({\cal I}\times{\cal R}_f)$, we have
\begin{equation}\label{Dc}
f(x,\varepsilon)={1\over
2^n}\sum_{i=1}^{2^{n-1}}f(x-{(2i-1)\over2^n}\varepsilon,{\varepsilon\over
2^n})+f(x+{(2i-1)\over2^n}\varepsilon,{\varepsilon\over
2^n}),\quad \forall n\geq1.
\end{equation}
\end{lemma}

\begin{proof} For n=1, the formula (\ref{Dc}) is verified (addition
property of integral) and we have
$f(x,\varepsilon)={1\over2}(f(x-{\varepsilon\over2},{\varepsilon\over2})+f(x+{\varepsilon\over2},{\varepsilon\over2}))$.
We suppose that we have the formula (\ref{Dc}) until the order n,
and let us prove it at the order n+1. We have $$
f(x,\varepsilon)={1\over
2^n}\sum_{i=1}^{2^{n-1}}f(x-{(2i-1)\over2^n}\varepsilon,{\varepsilon\over
2^n})+f(x+{(2i-1)\over2^n}\varepsilon,{\varepsilon\over 2^n})$$
and using the addition given by the formula (\ref{d0}) of the
functions
$$f(x-{(2i-1)\over2^n}\varepsilon,{\varepsilon\over
2^n})\qquad\hbox{ and}\qquad
f(x+{(2i-1)\over2^n}\varepsilon,{\varepsilon\over 2^n}),\quad
\hbox{we obtain}$$
$$f(x,\varepsilon)={1\over
2^{n+1}}\sum_{i=1}^{2^{n-1}}f(x-{(2i-1)\over2^n}\varepsilon-{\varepsilon\over
2^{n+1}},{\varepsilon\over
2^{n+1}})+f(x-{(2i-1)\over2^n}\varepsilon+{\varepsilon\over
2^{n+1}},{\varepsilon\over
2^{n+1}})$$$$+f(x+{(2i-1)\over2^n}\varepsilon-{\varepsilon\over
2^{n+1}},{\varepsilon\over
2^{n+1}})+f(x+{(2i-1)\over2^n}\varepsilon+{\varepsilon\over
2^{n+1}},{\varepsilon\over 2^{n+1}}),\quad \hbox{then}$$
$$f(x,\varepsilon)={1\over
2^{n+1}}\sum_{i=1}^{2^{n-1}}f(x-{(4i-1)\over2^{n+1}}\varepsilon,{\varepsilon\over
2^{n+1}})+f(x-{(4i-3)\over2^{n+1}}\varepsilon,{\varepsilon\over
2^{n+1}})$$$$+f(x+{(4i-3)\over2^{n+1}}\varepsilon,{\varepsilon\over
2^{n+1}})+f(x+{(4i-1)\over2^{n+1}}\varepsilon,{\varepsilon\over
2^{n+1}}).$$ With an adequate rearrangement, our sum can be
written as

$f(x,\varepsilon)={1\over
2^{n+1}}\sum_{i=1}^{2^{n-1}}f(x-{(4i-1)\over2^{n+1}}\varepsilon,{\varepsilon\over
2^{n+1}})+f(x+{(4i-1)\over2^{n+1}}\varepsilon,{\varepsilon\over
2^{n+1}})$$$+{1\over
2^{n+1}}\sum_{i=1}^{2^{n-1}}f(x+{(4i-3)\over2^{n+1}}\varepsilon,{\varepsilon\over
2^{n+1}})+f(x-{(4i-3)\over2^{n+1}}\varepsilon,{\varepsilon\over
2^{n+1}}).$$ For simplicity, we put $f(x,\varepsilon)=E+O$ with

\begin{equation}\label{S1}
E={1\over
2^{n+1}}\sum_{i=1}^{2^{n-1}}f(x-{(4i-1)\over2^{n+1}}\varepsilon,{\varepsilon\over
2^{n+1}})+f(x+{(4i-1)\over2^{n+1}}\varepsilon,{\varepsilon\over
2^{n+1}})
\end{equation}

\begin{equation}\label{S2}
O={1\over
2^{n+1}}\sum_{i=1}^{2^{n-1}}f(x+{(4i-3)\over2^{n+1}}\varepsilon,{\varepsilon\over
2^{n+1}})+f(x-{(4i-3)\over2^{n+1}}\varepsilon,{\varepsilon\over
2^{n+1}}).
\end{equation}
Now, if we put $j=2i$ for the formula (\ref{S1}), and $j=2i-1$ for
the formula (\ref{S2}), we obtain the following sums: a) Sum over
even numbers:
\begin{equation}\label{S3}
E={1\over
2^{n+1}}\sum_{j=2}^{2^{n}}f(x-{(2j-1)\over2^{n+1}}\varepsilon,{\varepsilon\over
2^{n+1}})+f(x+{(2j-1)\over2^{n+1}}\varepsilon,{\varepsilon\over
2^{n+1}})
\end{equation}
b) Sum over odd numbers:
\begin{equation}\label{S4}
O={1\over
2^{n+1}}\sum_{j=1}^{2^{n}-1}f(x+{(2j-1)\over2^{n+1}}\varepsilon,{\varepsilon\over
2^{n+1}})+f(x-{(2j-1)\over2^{n+1}}\varepsilon,{\varepsilon\over
2^{n+1}}).
\end{equation}
Finally we see that we have

$f(x,\varepsilon)=E+O$
\begin{equation}
={1\over
2^{n+1}}\sum_{i=1}^{2^{n}}f(x-{(2i-1)\over2^{n+1}}\varepsilon,{\varepsilon\over
2^{n+1}})+f(x+{(2i-1)\over2^{n+1}}\varepsilon,{\varepsilon\over
2^{n+1}}),
\end{equation}
and then the result $\forall n\geq1$.
\end{proof}

{\bf Remark} The simple addition given by the formula (\ref{d0})
can be repeated n times to obtain the formula (\ref{Dc}), where
$f(x-{(2i-1)\over2^n}\varepsilon,{\varepsilon\over 2^n})$ and
$f(x+{(2i-1)\over2^n}\varepsilon,{\varepsilon\over 2^n})$ for all
$n\geq1$, $i=1,2\dots, 2^{n-1}$, are the mean functions of $f$ at
the resolution ${\varepsilon\over 2^n}$. As an example of nowhere
differentiable function we mention the Weierstrass function given
by $W_\alpha(x)=\sum_{m=1}^{+\infty}q^{-\alpha m}\exp{(iq^mx)}$

For $n=1$, we obtain 2 identical graphs of approximation.

For $n=2$, we obtain $2^2$ identical graphs of approximation.

For $n=3$, we obtain $2^3$ identical graphs of approximation.

For a given integer $n$, we obtain $2^n$ multiple identical graphs
of approximation. If we write
$f(x+{\varepsilon\over2},{\varepsilon\over2})$ as an addition of
two identical mean functions, we will obtain an odd number of
non-identical graphs.

\subsection{New Characterization}
Let consider, for $\varepsilon\in{\cal R}_f$,
$\Gamma_\varepsilon=\{ (x,y)\in\rR^2\ /\ y=f(x,\varepsilon),\
x\in{\cal I}\}$ the graph associated to the mean function given by
(\ref{M}).

Let $\Gamma_f=\{ (x,y)\in\rR^2\ /\ y=f(x), \ x\in{\cal I}\}$ be
the graph associated to the nowhere differentiable function $f$.
Let $(\rR^2,\rho)$ be a metric space, and let us consider the
Hausdorff (\cite{Ha}) measure given by

$d_h(\Gamma_{ \varepsilon},\Gamma_f)=Max\Big[\sup_{A\in\Gamma_{
\varepsilon}}\inf_{B\in \Gamma_f}\rho(A,B),\sup_{A\in
\Gamma_f}\inf_{B\in\Gamma_{\varepsilon}}\rho(A,B)\Big].$

The mean function given by (\ref{M}) converges uniformly to the
nowhere differentiable function $f$. We say that the curve
$\Gamma_{\varepsilon}$ converges to the graph $\Gamma_f$ if
$d_h(\Gamma_{ \varepsilon},\Gamma_f)$ tends to 0 as $\varepsilon$
tends to 0, and that two curves $\Gamma_1$ and $\Gamma_2$ are
disjoint if $d_h(\Gamma_{1},\Gamma_2)>0$. We have the following
criterion:
\begin{lemma}\label{L3}
Let $f$ be a continuous function on $\cal I\subset\rR$, then

i) $f$ is nowhere differentiable on $\cal I$ $\Longleftrightarrow$
$\forall\ \varepsilon\in {\cal R}_f$, $
d_h(\Gamma_\varepsilon^-,\Gamma_\varepsilon^+)>0.$

 ii) $f$ is differentiable on $\cal I$ $\Longleftrightarrow$
$\exists\ \varepsilon\in {\cal R}_f$ such that $
d_h(\Gamma_\varepsilon^-,\Gamma_\varepsilon^+)=0. $
\end{lemma}

\begin{proof} For $$
d_h(\Gamma^+_\varepsilon,\Gamma^-_\varepsilon)=Max\Big[\sup_{A\in\Gamma^+_{
\varepsilon}}\inf_{B\in \Gamma^-_\varepsilon}\rho(A,B),\sup_{A\in
\Gamma^-_\varepsilon}\inf_{B\in\Gamma^+_{\varepsilon}}\rho(A,B)\Big],\quad
\hbox{where}$$ $A=(x,y)\in\Gamma^+_\varepsilon$,
$B=(x',y')\in\Gamma^-_\varepsilon$, and\quad
$\rho(A,B)=\sqrt{(x-x')^2+(y-y')^2}$$$=\sqrt{(x-x')^2+(f(x+{\varepsilon\over
2},{\varepsilon\over 2}) -f(x'-{\varepsilon\over
2},{\varepsilon\over 2}))^2}.$$ For $x'=x+\varepsilon$ we have
$f(x'-{\varepsilon\over 2},{\varepsilon\over
2})=f(x+\varepsilon-{\varepsilon\over 2},{\varepsilon\over
2})=f(x+{\varepsilon\over 2},{\varepsilon\over 2}),$ then we have
$\rho(A,B)=\sqrt{\varepsilon^2}=\varepsilon$ and
$d_h(\Gamma^+_\varepsilon,\Gamma^-_\varepsilon)=\varepsilon,$ and
following the definition of ${\cal R}_f$ we can conclude.
\end{proof}

The last property can be considered as a criterion of nowhere
differentiability for functions.

\section{Fractal Manifold}

We have seen that the geometry of the nowhere differentiable
functions could be determined by the study
 of the variations of mean functions (\ref{M}), we
have seen also that the nowhere differentiability can be
characterized by considering the superposition of the graph of the
forward-backward mean functions. In this section we want to
construct manifolds, which present locally an appearance of new
structure at different scales (using differentiable objects), and
we want to know if it is possible to construct a differentiable
geometry on it. We will call it "fractal manifold". This new
object will allow us to construct a new space. The graphs of the
family of mean functions
$\Big(f(x,\varepsilon)\Big)_{\varepsilon>0}$ could be considered
globally as a  fractal manifold of dimension 1 (the graphs have
appearance of new structure at different resolutions), however,
locally it doesn't present any new structure for different
resolutions since it is homeomorphic to $\rR$. The new idea is to
consider the mixture of the family of mean functions and the local
information given by the superposition of the forward-backward
mean function that characterizes the nowhere differentiable
functions.
\subsection{Introduction of the $\varepsilon$-Manifolds}

 Let us consider $f_1,f_2$ and $f_3$ three continuous and nowhere differentiable functions,
  defined on the open interval ${\cal I}$. The graphs of these functions are given by
$ \Gamma_{i}({\cal I})=\Big\lbrace\  (x,y)\in\rR^2/y=f_i(x),
x\in{\cal I}\ \Big\rbrace,\ i=1,2,3.$
 Let us consider the backward-forward mean functions given by

$f_i(x-{\varepsilon\over2},{\varepsilon\over2})=\di{1\over\varepsilon}\int^x_{x-\varepsilon}f_i(t)dt,
\qquad \hbox{and}\quad
f_i(x+{\varepsilon\over2},{\varepsilon\over2})=\di{1\over\varepsilon}\int_x^{x+\varepsilon}f_i(t)dt,$
for $i=1,2,3$. We denote respectively their associated graphs:
$\Gamma_{1,\varepsilon}^\sigma, \Gamma_{2,\varepsilon}^\sigma,
\Gamma_{3,\varepsilon}^\sigma$ where $ \sigma=\pm$ for the
forward-backward. We see that the graphs
$\Gamma_{1,\varepsilon}^\sigma, \Gamma_{2,\varepsilon}^\sigma,
\Gamma_{3,\varepsilon}^\sigma$ converge to the graphs $\Gamma_1,
\Gamma_2$ and $\Gamma_3$ as $\varepsilon$ tends to 0 in a
Hausdorff metric space $(\rR^2,d_h)$. By definition of the set
${\cal R}_f$, we have ${\cal R}_{f_1}={\cal R}_{f_2}={\cal
R}_{f_3}$. The set
$\Gamma_{1,\varepsilon}^\sigma\times\Gamma_{2,\varepsilon}^\sigma\times
\Gamma_{3,\varepsilon}^\sigma$ is included in $\rR^6$. Let us
consider the metric space $(\rR^6,d_h)$, then following the proof
of the lemma \ref{L3}, it's easy to find
\begin{equation}
d_h\Big(\Gamma_{1,\varepsilon}^-\times\Gamma_{2,\varepsilon}^-\times
\Gamma_{3,\varepsilon}^-\ ,\
\Gamma_{1,\varepsilon}^+\times\Gamma_{2,\varepsilon}^+\times
\Gamma_{3,\varepsilon}^+\Big)=\sqrt{3}\varepsilon.
\end{equation}
which means that the  Hausdorff distance between
$\Gamma_{1,\varepsilon}^\sigma\times\Gamma_{2,\varepsilon}^\sigma\times
\Gamma_{3,\varepsilon}^\sigma$, $\sigma=\pm$, increases or
decreases as a line with slope $\sqrt{3}$. We introduce what is
called $\varepsilon$-manifold or a local double space at the scale
$\varepsilon$:

\begin{definition}
Let  $\varepsilon$ be in  ${\cal R}_f$ and let us consider the
application

$$T_\varepsilon : \prod_{i=1}^{3}\Gamma_{i\varepsilon}^{+}\times \{\varepsilon\} \longrightarrow
\prod_{i=1}^{3}\Gamma_{i\varepsilon}^{-}\times
\{\varepsilon\}\quad\hbox{defined by}$$

 $T_\varepsilon
((a_1,b_1),(a_2,b_2),(a_3,b_3))=((a_1+\varepsilon,b_1),(a_2+\varepsilon,b_2),
(a_3+\varepsilon,b_3)),$ where $(a_i,b_i)\in
\Gamma_{i\varepsilon}^{+}$, that is to say\ $b_i=\di
f_i(a_i+{\varepsilon \over 2}, {\varepsilon \over 2})={1\over
\varepsilon}\di \int _{a_i}^{a_i+\varepsilon}f_i(t)dt$,
 for  $i=1,2,3$.

\end{definition}

\begin{proposition}
For all $\varepsilon \in {\cal R}_f$, we have the following
properties:

1) the application $T_\varepsilon$ is well defined.

2) $T_\varepsilon^{-1}$ exists.

3) $T_\varepsilon$ is an homeomorphism.

4) $T_\varepsilon $ and ${T_\varepsilon}^{-1}$ are differentiable.
\end{proposition}

\begin{proof} 1) Let $((a_1,b_1),(a_2,b_2),(a_3,b_3))\in
 \prod_{i=1}^{3}\Gamma_{i\varepsilon}^{+}\times \{\varepsilon\}$, then for
 $i=1,2,3,$ we have
$b_i=\di f_i(a_i+{\varepsilon \over 2}, {\varepsilon \over
2})=f_i((a_i+\varepsilon)-{\varepsilon \over 2}, {\varepsilon
\over 2},)$ then $(a_i+\varepsilon, b_i)\in
\Gamma_{i\varepsilon}^{-},$ and
$((a_1+\varepsilon,b_1),(a_2+\varepsilon,b_2),
(a_3+\varepsilon,b_3))\in
\prod_{i=1}^{3}\Gamma_{i\varepsilon}^{-}\times \{\varepsilon\}.$

2) $(T_\varepsilon)^{-1}
((a_1,b_1),(a_2,b_2),(a_3,b_3))=((a_1-\varepsilon,b_1),(a_2-\varepsilon,b_2),
(a_3-\varepsilon,b_3))$

3) and 4) $T_\varepsilon$ and $(T_\varepsilon)^{-1}$ are
continuous, differentiable as composite functions of continuous
and differentiable functions.
\end{proof}

\begin{definition}
Let $\varepsilon $ be in ${\cal R}_f$, and $M_\varepsilon$ be an
Hausdorff topological space. We say that $M_\varepsilon$ is an
$\varepsilon$-manifold if for every point $x \in M_\varepsilon$,
there exist a neighborhood $\Omega _{\varepsilon}$ of $x$ in
$M_\varepsilon$, a map $\varphi _\varepsilon$, and two open sets
$V^{+} _{\varepsilon}$ of
$\prod_{i=1}^{3}\Gamma_{i\varepsilon}^{+}\times \{\varepsilon\} $
and $V^{-} _{\varepsilon}$ of
$\prod_{i=1}^{3}\Gamma_{i\varepsilon}^{-}\times \{\varepsilon\} $
such that $\varphi _\varepsilon : \Omega _{\varepsilon}
\longrightarrow V^{+} _{\varepsilon}$, and $T_\varepsilon \circ
\varphi _\varepsilon: \Omega _{\varepsilon} \longrightarrow V^{-}
_{\varepsilon} $ are two homeomorphisms.
\end{definition}

 \unitlength=0.7cm
\begin{picture}(6,6)

\put(2,4.9){$M_\varepsilon$}


\put(3.6,4.45){\vector(2,-1){3.7}}

\put(4,4.8){\vector(2,0){3.}}

\put(7.8,4.5){\vector(0,-1){1.5}}


\put(5,5.2){$\varphi_\varepsilon$}

\put(9,3.6){$T_\varepsilon$}

\put(5.5,3.6){$T_\varepsilon \circ \varphi _\varepsilon$}


\put(7.5,4.8){$\prod_{i=1}^{3}\Gamma_{i\varepsilon}^{+}\times
\{\varepsilon\} $}

\put(7.5,2.5){$\prod_{i=1}^{3}\Gamma_{i\varepsilon}^{-}\times
\{\varepsilon\} $}

\put(5,0.8){\small Figure 5. Diagram of $\varepsilon$-manifold}
 \thicklines
\end{picture}

One can just say that $M_\varepsilon$ is an $\varepsilon$-manifold
if $M_\varepsilon$ is homeomorphic to
$\prod_{i=1}^{3}\Gamma_{i\varepsilon}^{+}\times \{\varepsilon\} $,
and we have automatically the second homeomorphism $T_\varepsilon
\circ \varphi _\varepsilon$.

\begin{definition}
A local chart on $M_\varepsilon$, for a given $\varepsilon \in
{\cal R}_f$, is a triplet $(\Omega_{\varepsilon}, \varphi
_\varepsilon, T_\varepsilon \circ \varphi _\varepsilon),$ where
$\Omega_{\varepsilon}$ is an open set of $M_\varepsilon$, $\varphi
_\varepsilon $ is an homeomorphism from $\Omega_{\varepsilon}$ to
an open set $V^+_\varepsilon$ of
$\prod_{i=1}^{3}\Gamma_{i\varepsilon}^{+}\times \{\varepsilon\} $
and $T_\varepsilon \circ \varphi _\varepsilon$ is an homeomorphism
from $\Omega_{\varepsilon}$ to an open set $V^-_\varepsilon$ of
$\prod_{i=1}^{3}\Gamma_{i\varepsilon}^{-}\times \{\varepsilon\} $.

A collection $(\Omega_{\varepsilon, i}, \varphi_{\varepsilon, i},
\psi _{\varepsilon,i})_{i\in J}$ of local charts such that $\cup
_{i\in J}\Omega _{\varepsilon,i}=M_\varepsilon$ is called an
atlas. The coordinates of $x\in \Omega_{\varepsilon}$ related to
the local chart $(\Omega_{\varepsilon}, \varphi _\varepsilon,
T_\varepsilon \circ \varphi _\varepsilon)$ are the coordinates of
the point $\varphi _\varepsilon (x)$ in
$\prod_{i=1}^{3}\Gamma_{i\varepsilon}^{+}\times \{\varepsilon\} $,
and of the point $T_\varepsilon \circ \varphi _\varepsilon(x)$ in
$\prod_{i=1}^{3}\Gamma_{i\varepsilon}^{-}\times \{\varepsilon\} $.
\end{definition}

\begin{definition}
An atlas of class ${\cal C}^1$, on $M_{\varepsilon}$ is an atlas
for which all changes of charts are ${\cal C}^1$. That is to say,
if $(\Omega_{\varepsilon,i},\varphi_{\varepsilon
,i},T_\varepsilon\circ\varphi_{\varepsilon ,i})$, and
$(\Omega_{\varepsilon,j},\varphi_{\varepsilon
,j},T_\varepsilon\circ\varphi_{\varepsilon ,j})$ are two local
charts
 with $\Omega_{\varepsilon,i}\cap\Omega_{\varepsilon,j}\not=\emptyset$,
 then the map change of charts
$\varphi_{\varepsilon, ij}=\varphi_{\varepsilon
,j}\circ(\varphi_{\varepsilon ,i})^{-1}$ from
$\varphi_{\varepsilon ,i}(\Omega_{\varepsilon,i}\cap
\Omega_{\varepsilon,j})$ to $\varphi_{\varepsilon
,j}(\Omega_{\varepsilon,i}\cap \Omega_{\varepsilon,j})$ is a
diffeomorphism of class ${\cal C}^1$, and the map change of charts
$\Big(T_\varepsilon\circ\varphi_{\varepsilon
,j}\Big)\circ\Big(T_\varepsilon\circ\varphi_{\varepsilon
,i}\Big)^{-1}$ from $T_\varepsilon\circ\varphi_{\varepsilon
,i}(\Omega_{\varepsilon,i}\cap \Omega_{\varepsilon,j})$ to
$T_\varepsilon\circ\varphi_{\varepsilon
,j}(\Omega_{\varepsilon,i}\cap \Omega_{\varepsilon,j})$ is a
diffeomorphism of class ${\cal C}^1$.
\end{definition}

We consider the following relation of equivalence between atlases
of class ${\cal C}^1$ on $M_{\varepsilon}$:  two atlases
$(\Omega_{\varepsilon,i},\varphi_{\varepsilon
,i},T_\varepsilon\circ\varphi_{\varepsilon ,i})_{i\in I}$, and
$(\Omega_{\varepsilon,j},\varphi_{\varepsilon,j},T_\varepsilon\circ\varphi_{\varepsilon,j})_{j\in
J}$ of class ${\cal C}^1$ are said to be equivalent if their union
is an atlas of class ${\cal C}^1$. That is to say that
$\varphi_{\varepsilon ,i}\circ(\varphi_{\varepsilon,j})^{-1}$ is
 ${\cal C}^1$ on
$\varphi_{\varepsilon,j}(\Omega_{\varepsilon,i}\cap
\Omega_{\varepsilon,j})$ and
$T_\varepsilon\circ\varphi_{\varepsilon,i}\circ(T_\varepsilon\circ\varphi_{\varepsilon,j})^{-1}$
is ${\cal C}^1$ on
$T_\varepsilon\circ\varphi_{\varepsilon,j}(\Omega_{\varepsilon,i}\cap
\Omega_{\varepsilon,j})$ when $\Omega_{\varepsilon,i}\cap
\Omega_{\varepsilon,j}\not=\emptyset$.

\begin{definition}
A differentiable $\varepsilon$-manifold of class ${\cal C}^1$,
$\varepsilon\in{\cal R}_f$, is an $\varepsilon$-manifold together
with an equivalence class ${\cal C}^1$ atlases.
\end{definition}

{\bf Examples of Atlas: } 1) The basic example is given by the
space $\prod_{i=1}^3\Gamma_{i,\varepsilon}^-$, for
$\varepsilon\in{\cal R}_f$. The set of one element $\lbrace\
\Big(\prod_{i=1}^3\Gamma_{i,\varepsilon}^-,id,T_\varepsilon^-\Big)\
\rbrace$ is an atlas of class ${\cal C}^1$, where

$$T_\varepsilon^-: \prod_{i=1}^3\Gamma_{i,\varepsilon}^-\longrightarrow
\prod_{i=1}^3\Gamma_{i,\varepsilon}^+$$
$$\Big((x_1,y_1),(x_2,y_2),(x_3,y_3)\Big)\longmapsto\Big((x_1-\varepsilon,y_1),
(x_2-\varepsilon,y_2),(x_3-\varepsilon,y_3) \Big),$$ and $id$ is
the identity of $\prod_{i=1}^3\Gamma_{i,\varepsilon}^-$.

2) Similarly, an example is given by the space
$\prod_{i=1}^3\Gamma_{i,\varepsilon}^+$, for
$\varepsilon\in\rR^{*+}$. The set of one element $\lbrace\
\Big(\prod_{i=1}^3\Gamma_{i,\varepsilon}^+,id,T_\varepsilon^+\Big)\
\rbrace$, where
$$
T_\varepsilon^+:\prod_{i=1}^3\Gamma_{i,\varepsilon}^+
\longrightarrow \prod_{i=1}^3\Gamma_{i,\varepsilon}^-$$$$
\Big((x_1,y_1),(x_2,y_2),(x_3,y_3)\Big)  \longmapsto
\Big((x_1+\varepsilon,y_1),(x_2+\varepsilon,y_2),
(x_3+\varepsilon,y_3)\Big)$$

and where $id$ is the identity of
$\prod_{i=1}^3\Gamma_{i,\varepsilon}^+$, is an atlas of class
${\cal C}^1$.

\subsection{Fractal Manifolds: Prototype}

\subsubsection{Diagonal topology}

\begin{definition}Let $A=\bigcup _{\varepsilon \in I}
A_\varepsilon$ and $B=\bigcup _{\varepsilon \in I} B_\varepsilon$
be two subsets of $E= \bigcup _{\varepsilon \in I} E_\varepsilon$
 union of Hausdorff topological spaces, $I\subset\rR^+$ an
interval, such that $\forall \varepsilon\in I$, $A_\varepsilon$
and $B_\varepsilon$ are subsets of $E_\varepsilon$. We call
diagonal intersection between $A$ and $B$ the set
\begin{equation}
A\ \widetilde{\cap}\ B=\bigcup _{\varepsilon \in
I}\Big(A_\varepsilon\cap B_\varepsilon\Big)
\end{equation}

\end{definition}

\begin{definition}
Let $I\subset\rR^+$ be an interval. A diagonal topology ${\cal
T}_d\subset{\cal P}(E)$ of a set $E=\bigcup _{\varepsilon \in I}
E_\varepsilon$ union of Hausdorff topological spaces, where the
 $E_\varepsilon$ are all disjoint or
all the same, consists of subsets of $E$ that verify the following
axioms:

(i) $\phi\in{\cal T}_d$, and $E\in{\cal T}_d$,

(ii) $\omega_1\in{\cal T}_d$, $\omega_2\in{\cal T}_d$\quad
$\Rightarrow$\quad $\omega_1\ \widetilde{\cap}\ \omega_2\in{\cal
T}_d$,

(iii) $\omega_i\in{\cal T}_d$, $\forall i\in J$\quad
$\Rightarrow$\quad $\di\bigcup_{i\in J}\omega_i\in{\cal T}_d$.

The elements $\omega_i\in{\cal T}_d$ are called open sets,
$(E,{\cal T}_d)$ is called diagonal topological space.
\end{definition}

\begin{proposition}
Let $E=\bigcup _{\varepsilon \in I} E_\varepsilon$ be an union of
Hausdorff topological spaces all disjoint or all the same, and let
$${\cal T}_d=\{\ \Omega=\cup_{\varepsilon\in
I}\Omega_\varepsilon\subset E/\quad\forall \varepsilon\in I, \
\Omega_\varepsilon\in E_{\varepsilon}\ \}.$$ Then ${\cal T}_d$ is
a diagonal topology on E.
\end{proposition}

\begin{proof} (i) is obvious.

(ii) If $\omega_1=\di\cup_{\varepsilon\in
I}\Omega_{1\varepsilon}\in{\cal T}_d$,\quad and
$\omega_2=\di\cup_{\varepsilon\in I}\Omega_{2\varepsilon}\in{\cal
T}_d$, then we have $$\omega_1\ \widetilde{\cap}\ \omega_2=\bigcup
_{\varepsilon \in I}\Big(\Omega_{1\varepsilon}\cap
\Omega_{2\varepsilon}\Big)\in{\cal T}_d.$$

(iii) For $\omega_i=\di\cup_{\varepsilon\in
I}\Omega_{i\varepsilon}\in{\cal T}_d$, $\forall i\in J$,\quad we
have $$\di\bigcup_{i\in J}\omega_i=\di\bigcup_{i\in
J}\cup_{\varepsilon\in {\cal
R}_f}\Omega_{i\varepsilon}=\di\bigcup_{\varepsilon\in{\cal R}_f}
\cup_{i\in J}\Omega_{i\varepsilon}$$ since $E_\varepsilon$ is a
topological space then we have $\di\bigcup_{i\in
J}\omega_i\in{\cal T}_d$
\end{proof}

\begin{definition}
Let $E=\bigcup _{\varepsilon \in I} E_\varepsilon$ be an union of
Hausdorff topological spaces all disjoint or all the same. Let
$x_\varepsilon\in E_\varepsilon$ $\forall \varepsilon\in I$. A
subset $\Omega\subset E$ is called a diagonal neighborhood of
$x_\varepsilon$, $\forall \varepsilon\in I$, if there exist
$\omega\in{\cal T}_d$ such that $\omega\subset\Omega$ and $\forall
\varepsilon\in I,$ $x_\varepsilon\in\omega$.
\end{definition}

\subsubsection{Prototype}

Let us consider a family of Hausdorff topological spaces
$E_\varepsilon,\  \forall \varepsilon \in I$, $I\subset\rR^+$, all
disjoint or all the same, and $E= \bigcup _{\varepsilon \in I}
E_\varepsilon$. Let $x:I\longrightarrow E$ be a continuous path on
$E$. If $\forall \varepsilon \in I$, $\Omega _\varepsilon$ is an
open neighborhood of $x(\varepsilon)$ in $E_\varepsilon$ , then
the set $\Omega(Range(x)) =\bigcup _ {\varepsilon \in I} \Omega
_\varepsilon$ is a diagonal neighborhood of the set
$Range(x)=\bigcup _{\varepsilon\in {I}}\{x(\varepsilon)\}$ in $E$.

\begin{definition}\label{IS}
Let $E=\bigcup _{\varepsilon \in I} E_\varepsilon$ union of
Hausdorff topological spaces, where the Hausdorff topological
spaces $E_\varepsilon$ are all disjoint or all the
same\footnote{Either $E=\cup_{\varepsilon\in I}E_\varepsilon$ is a
disjoint union, or $\forall\varepsilon\in I$, $E_\varepsilon=E_0$
and then $E=E_0$.}. We say that $E$ admits an internal structure
$x$ on $P\in E$, if there exists a ${\cal C}^0$ parametric path
\begin{equation}
\left .
\begin{array}{lll}
x: I & \longrightarrow & \cup_{\varepsilon\in I} E_{\varepsilon} \\
\varepsilon & \longmapsto & x(\varepsilon)\in E_\varepsilon ,
\end{array}
\right .
\end{equation}
 such that $\forall
\varepsilon\in I$, $Range (x) \cap
E_\varepsilon=\Big\{x(\varepsilon)\Big\}$, and $\exists\
\varepsilon'\in I$ such that $P=x(\varepsilon')\in
E_{\varepsilon'}$.
\end{definition}

{\bf Remark.} The continuity of the internal structure is defined
by the diagonal topology ${\cal T}_d$ (ie. $\forall\ \Omega\
\in{\cal T}_d$, $x^{-1}(\Omega)$ is an open set of $I$).

\begin{definition}
Let $E=\bigcup _{\varepsilon \in I} E_\varepsilon$ be an union of
Hausdorff topological spaces all disjoint or all the same. Let
$x$, $y$ be two internal structures on it. We say that $x\sim y$
$\Leftrightarrow$

 i) $\exists\ \varepsilon'\in{\cal R}_f$ such that $x(\varepsilon')=y(\varepsilon')$.

 ii) $\exists\ \theta: I\longrightarrow I$ diffeomorphism such that
$x=y\circ\theta$.
\end{definition}

\begin{proposition}\label{P1}
1) The relation $"\sim"$ is an equivalence relation.

2) If $x\sim y$ then $x=y$.
\end{proposition}

\begin{proof} 1) Using the definition of internal structure, it is not
difficult to see that $"\sim"$ is an equivalence relation.

2) Let us consider two paths of class ${\cal C}^0$ $x$ and $y$
such that $x\sim y$, we have then $\forall \varepsilon\in I$,
$x(\varepsilon)\in E_\varepsilon$ and $y(\varepsilon)\in
E_\varepsilon$, and we have
$x(\varepsilon)=y\circ\theta(\varepsilon)$, $\forall
\varepsilon\in I$. If we suppose that $\theta\not=id$, then there
exists $\varepsilon_0\in I$ such that
$\theta(\varepsilon_0)\not=\varepsilon_0$, we put
$\theta(\varepsilon_0)=\varepsilon'$. We have
$x(\varepsilon_0)=y\circ\theta(\varepsilon_0)$ then
$x(\varepsilon_0)=y(\varepsilon')$ which yields
$x(\varepsilon_0)\in E_{\varepsilon_0}$ and $x(\varepsilon_0)\in
E_{\varepsilon'}$, impossible then $\theta=id$ and
$\forall\varepsilon\in I$, $x(\varepsilon)=y(\varepsilon)$.
\end{proof}

\begin{definition}
Let $E=\bigcup _{\varepsilon \in I} E_\varepsilon$ be an union of
Hausdorff topological spaces all disjoint or all the same. Two
points of $E$ are equivalent if only if their internal structure
are equal.
\end{definition}

\begin{definition}
 Let $E=\bigcup _{\varepsilon \in I} E_\varepsilon$ be an union of
Hausdorff topological spaces all disjoint or all the same. Let
$x:I\subset\rR\longrightarrow\cup_{\varepsilon\in I}
E_{\varepsilon}$ be an internal structure on it. We call object of
$E$ the set $Range (x).$
\end{definition}

\begin{definition}\label{D2}
A diagonal topological space $(M,{\cal T}_d)$ is called fractal
manifold if $M=\bigcup _{\varepsilon \in {\cal
R}_f}M_\varepsilon$, where for all $\varepsilon \in {\cal R}_f$,
$M_\varepsilon$ is an $\varepsilon$-manifolds,
 and if $\forall P\in M$, $M$ admits an internal structure $x$ on $P$ such
that there exist a neighborhood
 $\Omega(Range(x))=\cup_{\varepsilon\in {\cal R}_f}\Omega_\varepsilon$,
with $\Omega_\varepsilon$ a neighborhood of $x(\varepsilon)$ in
$M_\varepsilon$, two open sets $V^+=\cup_{\varepsilon\in {\cal
R}_f} V_\varepsilon^+$ and $V^-=\cup_{\varepsilon\in {\cal R}_f}
V_\varepsilon^-$, where $V_\varepsilon^\sigma$ is an open set in
$\Pi_{i=1}^3\Gamma_{i\varepsilon}^\sigma\times\{\varepsilon\}$ for
$\sigma=\pm$, and there exist two families of maps
$(\varphi_\varepsilon)_{\varepsilon\in {\cal R}_f}$ and
$(T_\varepsilon\circ\varphi_\varepsilon)_{\varepsilon\in {\cal
R}_f}$ such that
$\varphi_\varepsilon:\Omega_\varepsilon\longrightarrow
V_\varepsilon^+ $ and
$T_\varepsilon\circ\varphi_\varepsilon:\Omega_\varepsilon\longrightarrow
V_\varepsilon^-$ are homeomorphisms for all ${\varepsilon\in {\cal
R}_f}$.
\end{definition}

{\bf Remark.} 1) Of course, if the family
$(\varphi_\varepsilon)_{\varepsilon\in {\cal R}_f}$ exists, then
the family
$(T_\varepsilon\circ\varphi_\varepsilon)_{\varepsilon\in {\cal
R}_f}$ exists automatically.

2) According to the Proposition \ref{P1}, we can associate for
each $P\in M$ only one path, and two points of $M$ are equivalent
if only if they are on the same path.

\begin{definition}
A local chart on fractal manifold $M$ is a triplet $(\Omega,
\varphi, T \circ \varphi)$, where $\Omega=\bigcup _{\varepsilon
\in {\cal R}_f}\Omega_{\varepsilon}$ is an open set of $M$,
$\varphi $ is a family of homeomorphisms $\varphi_\varepsilon$
from $\Omega_{\varepsilon}$ onto an open set $V^+_\varepsilon$ of
$\prod_{i=1}^{3}\Gamma_{i\varepsilon}^{+}\times \{\varepsilon\} $
and $T\circ \varphi$ is a family of homeomorphisms
$T_\varepsilon\circ \varphi_\varepsilon$ from
$\Omega_{\varepsilon}$ onto an open set $V^-_\varepsilon$ of
$\prod_{i=1}^{3}\Gamma_{i\varepsilon}^{-}\times \{\varepsilon\} $
for all $\varepsilon\in{\cal R}_f$. A collection $(\Omega_{i},
\varphi_{i}, (T\circ \varphi)_{i})_{i\in J}$ of local charts on
the fractal manifold $M$ such that $\cup _{i\in J}\Omega _i=\cup
_{i\in J}\bigcup _{\varepsilon \in {\cal
R}_f}\Omega_{i,\varepsilon}=\bigcup _{\varepsilon \in {\cal
R}_f}\cup _{i\in J}\Omega_{i,\varepsilon}=\bigcup _{\varepsilon
\in {\cal R}_f}M_\varepsilon=M,$ where $\cup _{i\in
J}\Omega_{i,\varepsilon}=M_\varepsilon$, is called an atlas. The
coordinates of an object $P\subset \Omega$ related to the local
chart $(\Omega, \varphi, T \circ \varphi )$ are the coordinates of
the object $\varphi (P)$ in $\bigcup _{\varepsilon \in {\cal
R}_f}\prod_{i=1}^{3}\Gamma_{i\varepsilon}^{+}\times
\{\varepsilon\} $, and of the object $T \circ \varphi(P)$ in
$\bigcup _{\varepsilon \in {\cal
R}_f}\prod_{i=1}^{3}\Gamma_{i\varepsilon}^{-}\times
\{\varepsilon\} $.
\end{definition}

\begin{definition}
An atlas of class ${\cal C}^1$ on a fractal manifold $M$ is an
atlas for which all families of changes of charts are ${\cal
C}^1$. That is to say, if we have an atlas of class ${\cal C}^1$
on $M_\varepsilon$ for all $\varepsilon \in {\cal R}_f$, then we
have a family of atlases of class ${\cal C}^1$, which gives us an
atlas of class ${\cal C}^1$ on $M$.
\end{definition}

We consider the following {\it equivalence relation} between
atlases of class ${\cal C}^1$ on a fractal manifold $M$:  two
atlases $(\Omega_{i},\varphi_{i},(T\circ\varphi)_{i})_{i\in I}$,
and $(\Omega_{j},\varphi_{j},(T\circ\varphi)_{j})_{j\in J}$ of
class ${\cal C}^1$ are said to be equivalent if their union is an
atlas of class ${\cal C}^1$ on $M$. That is to say that if we have
an equivalence relation between atlases of class ${\cal C}^1$ on
$M_\varepsilon$ for all $\varepsilon \in {\cal R}_f$, then we have
a family of equivalence relations, which gives us an equivalence
relation between atlases of class ${\cal C}^1$ on the fractal
manifold $M$.

\begin{definition}
\label{D3} A fractal manifold $M$ of class ${\cal C}^1$ is a
fractal manifold together with an equivalence class of ${\cal
C}^1$ atlases.
\end{definition}

{\bf Remark.} 1) The element $x(\varepsilon)\in M_\varepsilon$ has
$6$ local coordinates in a rectangular coordinate system
(considered as a subspace of $(\rR^6,d_h)$). In a local chart we
have: $\varphi _ \varepsilon (x(\varepsilon))= \Big
((x_1(\varepsilon),y_1(\varepsilon)),(x_2(\varepsilon),y_2(\varepsilon)),
(x_3(\varepsilon),y_3(\varepsilon))\Big )$, element of
$\prod_{i=1}^{3}\Gamma_{i\varepsilon}^{+} \times \{\varepsilon\}$,
and
$$T_\varepsilon \circ \varphi _ \varepsilon (x(\varepsilon))=
\Big((x_1(\varepsilon)+\varepsilon,y_1(\varepsilon)),(x_2(\varepsilon)+\varepsilon,y_2(\varepsilon)),
(x_3(\varepsilon)+\varepsilon,y_3(\varepsilon))\Big ),$$ element
of
$\prod_{i=1}^{3}\Gamma_{i\varepsilon}^{-}\times\{\varepsilon\}$,
where $y_i(\varepsilon)= f_i(x_{i}(\varepsilon)+{\varepsilon \over
2}, {\varepsilon \over 2}) \quad for \quad i=1,2,3$.

2) The sets
$\di(\prod_{i=1}^{3}\Gamma_{i\varepsilon}^{\sigma})_{\varepsilon
\in {\cal R}_f}$, for $\sigma =\pm$, are disjoint for the
Hausdorff distance, but not necessarily for another metric. Even
when the sets $
\prod_{i=1}^{3}\Gamma_{i\varepsilon}^{\sigma}\times
\{\varepsilon\}$, for $\sigma=\pm$, $\varepsilon \in {\cal R}_f$,
are always disjoint.

 \subsection{Example of Fractal Manifold}

An example of fractal manifold is given by the following lemma and
we will see later on why this name.
\begin{lemma}\label{L2}
Let $f_1$,$f_2$, and $f_3$ be nowhere differentiable functions.
Let $f_i(x+{\varepsilon\over2},{\varepsilon\over2})$, and
$f_i(x+{3\varepsilon\over4},{\varepsilon\over4})$ be  mean
functions of the functions $f_i$, for i=1,2,3. If we associate the
graph $\Gamma_{i\varepsilon}^+$ to the function
$f_i(x+{\varepsilon\over2},{\varepsilon\over2})$ (respectively,
$\Gamma_{i{\varepsilon\over2}}^+$ to the function
$f_i(x+{3\varepsilon\over4},{\varepsilon\over4})$) for i=1,2,3.
Then  $\bigcup_{\varepsilon\in{\cal R}_f}
\prod_{i=1}^3\Gamma_{i\varepsilon}^+\times\{\varepsilon\}$
 is a ${\cal C}^1$ scale manifold
homeomorphic to $\bigcup_{\varepsilon\in{\cal R}_f}
\prod_{i=1}^3\Gamma_{i{\varepsilon\over2}}^+\times\{\varepsilon\}$,
 and we have the following diagram:
\end{lemma}

\unitlength=0.7cm
\begin{picture}(6,6)

\put(0,5.9){$M=\di\bigcup _{\varepsilon \in {\cal
R}_f}\prod_{i=1}^3\Gamma_{i\varepsilon}^+\times\{\varepsilon\}$}


\put(5.1,5.45){\vector(2,-1){3.7}}

\put(5.5,5.8){\vector(2,0){3.}}

\put(9.3,5.5){\vector(0,-1){1.5}}


\put(6.5,6.2){$(\varphi_\varepsilon)_\varepsilon$}

\put(10.5,4.6){$(T_{\varepsilon\over2})_\varepsilon$}

\put(4,4.){$(T_{\varepsilon\over2} \circ \varphi
_\varepsilon)_\varepsilon$}


\put(9,5.8){$\bigcup _{\varepsilon \in {\cal
R}_f}\prod_{i=1}^{3}\Gamma_{i{\varepsilon\over2}}^{+}\times
\{\varepsilon\} $}

\put(9,3){$\bigcup _{\varepsilon \in {\cal
R}_f}\prod_{i=1}^{3}\Gamma_{i{\varepsilon\over2}}^{-}\times
\{\varepsilon\}. $}

\put(2,0.8){\small Figure 6. Diagram of the scale manifold
$\bigcup_{\varepsilon\in{\cal R}_f}
\prod_{i=1}^3\Gamma_{i\varepsilon}^+\times\{\varepsilon\}$}

 \thicklines
\end{picture}

\begin{proof}
 If for $i=1,2,3$, $\Gamma_{i\varepsilon^+}$ represents the graph of the forward mean
function $f_i(x+{\varepsilon\over2},{\varepsilon\over2})$ of the
nowhere differentiable function $f_i$, at $\varepsilon$
resolution, and $\Gamma_{i\varepsilon\over2}^+$ represents the
graph of the forward mean function
$f_i(x+{3\varepsilon\over4},{\varepsilon\over4})$ of $f_i$ at
$\varepsilon\over2$ resolution, then one can define
$$\begin {array}{l} \varphi_\varepsilon :
\Gamma_{i\varepsilon^+} \longrightarrow  \Gamma_{i\varepsilon\over2}^{+}\\
\qquad (x,f_i(x+{\varepsilon\over2},{\varepsilon\over2}))
\longmapsto
(x+{\varepsilon\over4},f_i(x+{3\varepsilon\over4},{\varepsilon\over4})
).
\end {array}$$
Its not difficult to see that:

 1) $\varphi_\varepsilon$ is continuous: each coordinate function
 is continuous, as composite function of continuous functions.

 2) $\varphi^{-1}_\varepsilon$ exists:
$\varphi^{-1}_\varepsilon(x,y_i(x,\varepsilon))=(x-{\varepsilon\over4},y_i(x-{\varepsilon\over4},2\varepsilon))$

3) $\varphi^{-1}_\varepsilon$ is continuous for the same reason as
$\varphi_\varepsilon$.

Then $\varphi_\varepsilon$ is an homeomorphism from
$\Gamma_{i\varepsilon^+}$ onto $\Gamma_{i\varepsilon\over2}^{+}$,
for i=1,2,3. We can generalize this result to the product of three
graphs, and we obtain an homeomorphism of the form $
\varphi_\varepsilon :
\prod_{i=1}^3\Gamma_{i\varepsilon}^+\times\{\varepsilon\}
\longrightarrow
\prod_{i=1}^3\Gamma_{i\varepsilon\over2}^{+}\times\{\varepsilon\},
$ and we have the following diagram:

\unitlength=0.7cm
\begin{picture}(6,6)

\put(0,4.9){$M_\varepsilon=\prod_{i=1}^3\Gamma_{i\varepsilon}^+\times\{\varepsilon\}$}


\put(4.6,4.45){\vector(2,-1){3.7}}

\put(5,4.8){\vector(2,0){3.}}

\put(8.8,4.5){\vector(0,-1){1.5}}


\put(6,5.2){$\varphi_\varepsilon$}

\put(10,3.6){$T_{\varepsilon\over2}$}

\put(3.5,3.){$T_{\varepsilon\over2} \circ \varphi _\varepsilon$}


\put(8.5,4.8){$\prod_{i=1}^{3}\Gamma_{i{\varepsilon\over2}}^{+}\times
\{\varepsilon\} $}

\put(8.5,2){$\prod_{i=1}^{3}\Gamma_{i{\varepsilon\over2}}^{-}\times
\{\varepsilon\} .$}

\put(4.5,0.8){\small Figure 7. Diagram of
$\varepsilon\over2$-manifold}
 \thicklines
\end{picture}

Internal structure can be found on $\di\bigcup _{\varepsilon \in
{\cal
R}_f}\prod_{i=1}^3\Gamma_{i\varepsilon}^+\times\{\varepsilon\}$.

Indeed, $\forall P\in\di\bigcup _{\varepsilon \in {\cal
R}_f}\prod_{i=1}^3\Gamma_{i\varepsilon}^+\times\{\varepsilon\}$,
there exists $\varepsilon'\in{\cal R}_f$ such that
$P=x(\varepsilon')=\Big(x_1,y(x_1,\varepsilon'),x_2,y(x_2,\varepsilon'),x_3,y(x_3,\varepsilon')\Big)$
where
$y(x_i,\varepsilon')=f_i(x_i+{\varepsilon'\over2},{\varepsilon'\over2})$
for $i=1,2,3$, and where the internal structure is given by the
${\cal C}^0$ parametric path:
\begin{equation}
\begin{array}{l} x: {\cal R}_f  \longrightarrow
 \cup_{\varepsilon\in {\cal
R}_f}\prod_{i=1}^3\Gamma_{i\varepsilon}^+\times\{\varepsilon\} \\
 \varepsilon
\longmapsto
x(\varepsilon)=\Big(x_1,y(x_1,\varepsilon),x_2,y(x_2,\varepsilon),x_3,y(x_3,\varepsilon)\Big)\in
\prod_{i=1}^3\Gamma_{i\varepsilon}^+\times\{\varepsilon\},
\end{array}
\end{equation}
and we have $\forall\varepsilon\in {\cal R}_f$,
$Range(x)\cap\prod_{i=1}^3\Gamma_{i\varepsilon}^+\times\{\varepsilon\}=\Big\{x(\varepsilon)\Big\}.$
The $x_i$ are constant and the $y(x_i,\varepsilon)$ are of class
${\cal C}^1$. Using the definition \ref{D2}, we obtain a fractal
manifold
\begin{equation}
M=\di\bigcup _{\varepsilon \in {\cal R}_f}M_\varepsilon=\di\bigcup
_{\varepsilon \in {\cal
R}_f}\prod_{i=1}^3\Gamma_{i\varepsilon}^+\times\{\varepsilon\}.
\end{equation}
\end{proof}

\subsection{Elements of the Fractal Manifold}

 By definitions \ref{IS}, the map $x: {\cal R}_f \longrightarrow
M$ describes the evolution of one representative element
$x(\varepsilon)$ of $x$. This map is a continuous path of
$\varepsilon$, and objects of $M$ are not "points" but ranges of
differentiable paths parameterized by $\varepsilon\in{\cal R}_f$.
An element $x(\varepsilon)$ of $M_{\varepsilon}$ is represented in
local coordinates by two points, then an object $P$ of $M$ is
represented in local coordinates by $ Range(x^+)\cup Range(x^-)$
where the paths $x^+$ and $x^-$ are given by:

\begin{equation}\label{O1}
\left .
\begin{array}{lll}
x^{+} : {\cal R}_f  & \longrightarrow   \di\bigcup _{\varepsilon
\in {\cal R}_f} \prod_{i=1}^{3}\Gamma_{i\varepsilon}^{+}\times\{\varepsilon\} &\\
 &\varepsilon \longmapsto\varphi _{\varepsilon}(x(\varepsilon))&,
\end{array}
\right .
\end{equation}

and
\begin{equation}\label{O2}
\left .
\begin{array}{lll}
x^{-} : {\cal R}_f  & \longrightarrow   \di\bigcup _{\varepsilon
\in {\cal R}_f} \prod_{i=1}^{3}\Gamma_{i\varepsilon}^{-}\times\{\varepsilon\} &\\
 &\varepsilon \longmapsto T_{\varepsilon}\circ \varphi
_{\varepsilon}(x(\varepsilon))&.
\end{array}
\right .
\end{equation}

Then locally, an object of the fractal manifold $M$ is not a
"point" but a disjoint union of sets of points (double copy), it
can be seen as a double string.
\begin{equation}
\Big\backslash \Big/
\end{equation}

{\bf Remark.} We have $d(x^+(\varepsilon),
x^-(\varepsilon))=\varepsilon\sqrt{3}$ for all $\varepsilon \in
{\cal R}_f$ then the ranges of the two paths become closer as
$\varepsilon$ tends to 0, but $\varepsilon \not=0$. If an object
of fractal manifold is always locally represented by a double
string it will never deserve the name fractal. In the following
section we will prove the appearance of substructure into it.

\section{Process Mean of the Mean}

Actually, the principle of approximation of the nowhere
differentiable function by an infinity of forward and backward
mean functions, can be made for differentiable functions too.
Which means that we can define our nowhere differentiable function
by a family of differentiable functions, and we can also define
the family of differentiable functions by another family of
differentiable functions. The procedure can be repeated
indefinitely. Visibly, repeating the procedure seems to be
useless. In this part we will explore the hidden information we
get by repeating the procedure indefinitely. But first of all, we
have to introduce "points" in our new manifold.

\subsection{Points and Elements: Toward a New Geometry}

All the classical physical spaces are manifolds, sets of points
with structures. In physics, a point is a pyridoxal object with
zero extension, it's a limit entity that no one have ever seen.
Points are without depth and their accumulation gives a space or a
continuous manifold. Point is the origin of many difficulties in
physics: divergence in classical and quantum physics, singularity
in cosmology and general relativity, divergence in quantum field
theory....etc. The Heisenberg's uncertainly principle prohibits a
perfect localization, which means that point is inaccessible. Many
physicists tried to introduce a new space without points, with
elementary cell " Atom Space", with the hope to make disappear the
divergences and to clarify naturally the impossible localization
in quantum mechanics. In mathematics, point is assumed to be
dimensionless, in axiomatic geometry, usually a completely
undefined\footnote{In Euclid's geometry: a point is that which has
no parts \cite{HA,CA}.}(primitive) element, although there are
axiomatizations of geometry in which those properties of {\it
point} that are desired, are given by postulates{\footnote{In a
geometry in which line is a primitive element, points may be
defined as classes of lines that conform to certain requirements
(postulates). In n-dimensional metric analytic point geometry, a
point may be defined as an ordered n-tuple of numbers...etc. }}.
Any physical representation of a point must be of some size, if we
draw a dot (point) and we magnify it, we will find a small
surface. At this level, if we erase the small surface and we put
another dot instead, and we magnify it again, we will obtain
another small surface etc. We can repeat this procedure
indefinitely, and if we want to introduce a notion of point, we
have to take into account all the points of view given by
magnification. An intuitive new postulate is as follow: {\it A
point is that which has no parts for a given scale.}

\begin{lemma}\label{G}
Let $g_1$,$g_2$, and $g_3$ be differentiable functions, and let
$g_i(x+{\delta\over2},{\delta\over2})$, be the mean functions of
the functions $g_i$, for i=1,2,3. If we associate the graph
$\Gamma_{i0}$ to the functions $g_i(x)$ (respectively,
$\Gamma_{i{\varepsilon}}^\sigma$ to the functions
$g_i(x+\sigma{\varepsilon\over2},{\varepsilon\over2})$,
$\sigma=\mp$, for i=1,2,3). The product $\prod_{i=1}^3\Gamma_{i0}$
is a fractal manifold of class ${\cal C}^1$ homeomorphic to $
\bigcup_{\varepsilon\in{\cal R}_f}
\prod_{i=1}^3\Gamma_{i{\delta}}^+\times\{\delta\},$ and we have
the following diagram:
\end{lemma}

\unitlength=0.7cm
\begin{picture}(6,6)

\put(1,4.9){$M=\prod_{i=1}^3\Gamma_{i0}$}


\put(5.1,4.45){\vector(2,-1){3.7}}

\put(5.5,4.8){\vector(2,0){3.}}

\put(9.3,4.5){\vector(0,-1){1.5}}


\put(6.5,5.2){$(\varphi_\delta)_\delta$}

\put(10.5,3.6){$(T_{\delta})_\delta$}

\put(4,3.){$(T_{\delta} \circ \varphi _\delta)_\delta$}


\put(9,4.8){$\bigcup _{\delta \in {\cal
R}_g}\prod_{i=1}^{3}\Gamma_{i{\delta}}^{+}\times \{\delta\} $}

\put(9,2){$\bigcup _{\delta \in {\cal
R}_g}\prod_{i=1}^{3}\Gamma_{i{\delta}}^{-}\times \{\delta\}. $}

 \thicklines
\end{picture}
\begin{proof} Let $g$ be a differentiable function, and let
$\delta\in{\cal R}_g$. If we denote $\Gamma_0$ the graph of the
function $g$ and $\Gamma_\delta^+$ the graph of the forward mean
function $g(x+{\delta\over2},{\delta\over2})$ of $g$ at $\delta$
resolution. We consider  $ \varphi_\delta : \Gamma_0
\longrightarrow \Gamma_\delta^+$ defined by $(x,g(x,0))
\longmapsto
\di(x+{\delta\over2},g(x+{\delta\over2},{\delta\over2})),$ where
$g(x+{\delta\over2},{\delta\over2})=\di{1\over\delta}\int_x^{x+\delta}
g(t)dt$, and $g(x,0)=g(x)$. Then we have:

 1) $\varphi_\delta$ is continuous: each coordinate function
 is continuous, as composite function of continuous functions.

 2) $\varphi^{-1}_\delta$ exists:
$\varphi^{-1}_\delta(x,y(x,\delta))=(x-{\delta\over2},y(x-{\delta\over2}))$

3) $\varphi^{-1}_\delta$ is continuous for the same reason as
$\varphi_\delta$.

Then $\varphi_\delta$ is an homeomorphism from $\Gamma_0$ onto
$\Gamma_{\delta}^{+}\times\{\delta\}$. For three dimensions we
have to consider $\Gamma_{i{\delta}}^{+}$ the graph associated to
the forward mean functions of the differentiable functions $g_i$,
and we denote $\Gamma_{i0}$ the graph associated to the functions
$g_i$, for $i=1,2,3$. By the same, we obtain an homeomorphism
$\varphi_\delta : \prod_{i=1}^3\Gamma_{i0} \longrightarrow
\prod_{i=1}^3\Gamma_{i{\delta}}^{+}\times\{\delta\}$, and then
$\prod_{i=1}^3\Gamma_{i0}$ is a $\delta$-manifold, and we have the
diagram given by Fig.8.

\unitlength=0.7cm
\begin{picture}(6,6)

\put(1,4.9){$M_\delta=\prod_{i=1}^3\Gamma_{i0}$}


\put(5.1,4.45){\vector(2,-1){3.7}}

\put(5.5,4.8){\vector(2,0){3.}}

\put(9.3,4.5){\vector(0,-1){1.5}}


\put(6.5,5.2){$\varphi_\delta$}

\put(10.5,3.6){$T_{\delta}$}

\put(4,3.){$T_{\delta} \circ \varphi _\delta$}


\put(9,4.8){$\prod_{i=1}^{3}\Gamma_{i{\delta}}^{+}\times
\{\delta\} $}

\put(9,2){$\prod_{i=1}^{3}\Gamma_{i{\delta}}^{-}\times \{\delta\}
.$}

\put(4.5,0.8){\small Figure 8. Diagram of $\delta$-manifold}
 \thicklines
\end{picture}

 An internal structure can be found on $\di\bigcup _{\delta \in
{\cal R}_g}\prod_{i=1}^3\Gamma_{i0}=\prod_{i=1}^3\Gamma_{i0}$.
Indeed, for all $P\in\di\bigcup _{\varepsilon \in {\cal
R}_g}\prod_{i=1}^3\Gamma_{i0}$, we have
$P=\Big(x_1,y(x_1),x_2,y(x_2),x_3,y(x_3)\Big)$ and then for
$y(x_i)=y_i$ constant for $i=1,2,3$, $P=(x_1,y_1,x_2,y_2,x_3,y_3)$
(which means that for every scale we consider the same point), and
where a ${\cal C}^0$ parametric path $x$ is given by
\begin{equation}
\begin{array}{l} x: {\cal R}_g  \longrightarrow
 \cup_{\delta\in {\cal
R}_g}\prod_{i=1}^3\Gamma_{i0} \\
\delta \longmapsto x(\delta)=(x_1,y_1,x_2,y_2,x_3,y_3)\in
\prod_{i=1}^3\Gamma_{i0}
\end{array}
\end{equation}
and we have $\forall\delta\in {\cal R}_g$,
$Range(x)\cap\prod_{i=1}^3\Gamma_{i0}=\Big\{P\Big\}.$ Using the
definition \ref{D2}, for $M_\delta=\prod_{i=1}^3\Gamma_{i0}$, we
obtain a fractal manifold $ M=\di\bigcup _{\delta \in {\cal
R}_g}M_\delta=\di\bigcup _{\delta \in {\cal
R}_g}\prod_{i=1}^3\Gamma_{i0}=\prod_{i=1}^3\Gamma_{i0}, $
homeomorphic to $\di\bigcup _{\delta \in {\cal
R}_g}\prod_{i=1}^3\Gamma_{i{\delta}}^{+}\times\{\delta\}$, and
then we have the result. Since $\prod_{i=1}^3\Gamma_{i0}$ is
homeomorphic to $\prod_{i=1}^3\Gamma_{i0}\times\{\varepsilon\}$
for $\varepsilon\in{\cal R}_g$, then
$\prod_{i=1}^3\Gamma_{i0}\times\{\varepsilon\}$ is homeomorphic to
$\di\bigcup _{\delta \in {\cal
R}_g}\prod_{i=1}^3\Gamma_{i{\delta}}^{+}\times\{\delta\}\times\{\varepsilon\}$.
\end{proof}

\begin{corollary}\label{Cor1}
For $\varepsilon\in{\cal R}_g$, the manifold given by
$\prod_{i=1}^3\Gamma_{i0}\times\{\varepsilon\}$ is a fractal
manifold of class ${\cal C}^1$ homeomorphic to $\di\bigcup
_{\delta \in {\cal
R}_g}\prod_{i=1}^3\Gamma_{i{\delta}}^{+}\times\{\delta\}\times\{\varepsilon\}$,
and we have the following diagram:
\end{corollary}

\unitlength=0.7cm
\begin{picture}(6,6)

\put(1,4.9){$M=\prod_{i=1}^3\Gamma_{i0}\times\{\varepsilon\}$}


\put(5.1,4.45){\vector(2,-1){3.7}}

\put(5.5,4.8){\vector(2,0){3.}}

\put(9.3,4.5){\vector(0,-1){1.5}}


\put(6.5,5.2){$(\varphi_\delta)_{\delta,\varepsilon}$}

\put(10.5,3.6){$(T_{\delta})_{\delta,\varepsilon}$}

\put(4,3.){$(T_{\delta} \circ \varphi
_\delta)_{\delta,\varepsilon}$}


\put(9,4.8){$\bigcup _{\delta \in {\cal
R}_g}\prod_{i=1}^{3}\Gamma_{i{\delta}}^{+}\times
\{\delta\}\times\{\varepsilon\} $}

\put(9,2){$\bigcup _{\delta \in {\cal
R}_g}\prod_{i=1}^{3}\Gamma_{i{\delta}}^{-}\times
\{\delta\}\times\{\varepsilon\} .$}

\put(3.5,0.8){\small Figure 9. Diagram of $\delta$-manifold at a
given $\varepsilon$}
 \thicklines
\end{picture}

We can generalize the last result to a three dimensional
differentiable manifold $M_0$, indeed, if any three dimensional
differentiable manifold $M_0$ is homeomorphic to
$\prod_{i=1}^3\Gamma_{i0}$ (Lemma \ref{G}) then $M_0$ is a fractal
manifold
\begin{corollary}\label{Cor2}
Let $g_1$,$g_2$, and $g_3$ be differentiable functions, and let
$g_i(x+{\delta\over2},{\delta\over2})$, be the mean functions of
the functions $g_i$, for i=1,2,3. We associate the graph
$\Gamma_{i0}$ to the functions $g_i(x)$. If $M_0$ is a three
dimensional differentiable manifold homeomorphic to the product
$\prod_{i=1}^3\Gamma_{i0}$ then $M_0$ is a fractal manifold.
\end{corollary}

If we consider an object $P$ of the fractal manifold
$\prod_{i=1}^3\Gamma_{i0}\times\{\varepsilon\}$, it's represented
in local coordinates, in
$\prod_{i=1}^{3}\Gamma_{i{\delta}}^{\sigma}\times
\{\delta\}\times\{\varepsilon\}$, by two points denoted by
$x_\varepsilon^+(\delta)$, $x_\varepsilon^-(\delta)$ for all
$\delta\not=0$, and by one point for $\delta=0$ given by
$x_\varepsilon^+(0)=x_\varepsilon^-(0)$ (using a classical
notation of point). At this level, an object $P$ of
$\prod_{i=1}^3\Gamma_{i0}\times\{\varepsilon\}$ is represented in
local coordinates, in $\bigcup _{\delta \in {\cal
R}_g}\prod_{i=1}^{3}\Gamma_{i{\delta}}^{+}\times
\{\delta\}\times\{\varepsilon\} $, by $Range(x_\varepsilon^+)\cup
Range(x_\varepsilon^-)$, where the two paths $x_\varepsilon^+$ and
$x_\varepsilon^-$ are given by:

\begin{equation}
\left .
\begin{array}{lll}
x_\varepsilon^{+} : {\cal R}_g  & \longrightarrow   \di\bigcup
_{\delta
\in {\cal R}_g} \prod_{i=1}^{3}\Gamma_{i{\delta}}^{+}\times\{\delta\} &\\
 &\delta \longmapsto\varphi _{\delta}(x_\varepsilon(\delta))&,
\end{array}
\right .
\end{equation}

and
\begin{equation}
\left .
\begin{array}{lll}
x_\varepsilon^{-} : {\cal R}_g  & \longrightarrow   \di\bigcup
_{\delta
\in {\cal R}_g} \prod_{i=1}^{3}\Gamma_{i{\delta}}^{-}\times\{\delta\} &\\
 &\delta \longmapsto T_{{\delta}}\circ \varphi
_{\delta}(x_\varepsilon(\delta))&,
\end{array}
\right .
\end{equation}

 with $x_\varepsilon^{+}(0)=x_\varepsilon^{-}(0)$, then an object of the fractal manifold
 $M=\prod_{i=1}^3\Gamma_{i0}$ at the resolution $\varepsilon$
looks like
\begin{equation}\label{E1}
\bigvee
\end{equation}
 and we have $\di d(x_\varepsilon^+(\delta),
x_\varepsilon^-(\delta))=\sqrt{3}{\delta}$\quad for all $\delta
\in {\cal R}_g$, then the two paths become closer when $\delta$
tends to 0, and for $\delta=0$ the two strings intersect at one
point. An element $x(\varepsilon)$ of $M_{\varepsilon}$ introduced
in definition \ref{D2} is then represented in local coordinates by
two elements, one from
$\prod_{i=1}^{3}\Gamma_{i\varepsilon}^{+}\times \{\varepsilon\} $
which has the form given by (\ref{E1}), and one from
$\prod_{i=1}^{3}\Gamma_{i\varepsilon}^{-}\times \{\varepsilon\} $
which has also the form given by (\ref{E1})\footnote{We apply
corollary \ref{Cor1} for the graphs
$\Gamma_{i0}=\Gamma_{i\varepsilon}^+$ of
$g_i=f_i(x+{\varepsilon\over2},{\varepsilon\over2})$ for
i=1,2,3.}, and where the distance in a Hausdorff metric space
between the two elements is $\sqrt{3}\varepsilon$. Then an object
$P$ of $M$ is represented in local coordinates by $Range(x^+)\cup
Range(x^-)$ where the two paths $x^+$ and $x^-$ are given by the
formula (\ref{O1}) and (\ref{O2}) respectively, such that
$x^+(\varepsilon)$ and $x^-(\varepsilon)$ are two copies given by
the form (\ref{E1}):
\begin{equation}\label{E2}
\bigvee\quad\bigvee
\end{equation}
 At the end an object of  the fractal manifold $M=\bigcup _{\varepsilon \in {\cal
R}_f}M_\varepsilon$ is represented by the union of the range of
$x^+$ and $x^-$ given by the formula (\ref{O1}) and (\ref{O2})
respectively (Fig.10a).

\unitlength=1cm
\begin{picture}(7,7)
 \thicklines
\put(5,3.4){\drawline(1,0.4)(1.12,.6)(1.4,1.8)(1.3,1.66)(1.28,1.84)(1,.6)(.72,1.84)(.7,1.66)(.6,1.8)(.88,.6)(1,.4)}
\put(1.5,3.4){\drawline(1.2,1)(1,.4)(.8,1)} \put(0,3.8){$\cdot$}


\put(8.54,5.2){\drawline(0.45,0.1)(0.52,0.17)(0.59,0.5)(0.52,0.43)(0.49,0.53)(0.425,0.2)(0.2,0.47)(0.22,0.37)(0.12,0.4)(0.35,0.14)(0.45,0.1)}
\put(8.585,5.05){\drawline(0.45,0.1)(0.52,0.17)(0.59,0.5)(0.52,0.43)(0.49,0.53)(0.425,0.2)(0.2,0.47)(0.22,0.37)(0.12,0.4)(0.35,0.14)(0.45,0.1)}
\put(8.63,4.9){\drawline(0.45,0.1)(0.52,0.17)(0.59,0.5)(0.52,0.43)(0.49,0.53)(0.425,0.2)(0.2,0.47)(0.22,0.37)(0.12,0.4)(0.35,0.14)(0.45,0.1)}
\put(8.675,4.75){\drawline(0.45,0.1)(0.52,0.17)(0.59,0.5)(0.52,0.43)(0.49,0.53)(0.425,0.2)(0.2,0.47)(0.22,0.37)(0.12,0.4)(0.35,0.14)(0.45,0.1)}
\put(8.72,4.6){\drawline(0.45,0.1)(0.52,0.17)(0.59,0.5)(0.52,0.43)(0.49,0.53)(0.425,0.2)(0.2,0.47)(0.22,0.37)(0.12,0.4)(0.35,0.14)(0.45,0.1)}
\put(8.765,4.45){\drawline(0.45,0.1)(0.52,0.17)(0.59,0.5)(0.52,0.43)(0.49,0.53)(0.425,0.2)(0.2,0.47)(0.22,0.37)(0.12,0.4)(0.35,0.14)(0.45,0.1)}
\put(8.81,4.3){\drawline(0.45,0.1)(0.52,0.17)(0.59,0.5)(0.52,0.43)(0.49,0.53)(0.425,0.2)(0.2,0.47)(0.22,0.37)(0.12,0.4)(0.35,0.14)(0.45,0.1)}
\put(8.855,4.15){\drawline(0.45,0.1)(0.52,0.17)(0.59,0.5)(0.52,0.43)(0.49,0.53)(0.425,0.2)(0.2,0.47)(0.22,0.37)(0.12,0.4)(0.35,0.14)(0.45,0.1)}
\put(8.9,4){\drawline(0.45,0.1)(0.52,0.17)(0.59,0.5)(0.52,0.43)(0.49,0.53)(0.425,0.2)(0.2,0.47)(0.22,0.37)(0.12,0.4)(0.35,0.14)(0.45,0.1)}
\put(8.945,3.85){\drawline(0.45,0.1)(0.52,0.17)(0.59,0.5)(0.52,0.43)(0.49,0.53)(0.425,0.2)(0.2,0.47)(0.22,0.37)(0.12,0.4)(0.35,0.14)(0.45,0.1)}
\put(8.99,3.7){\drawline(0.45,0.1)(0.52,0.17)(0.59,0.5)(0.52,0.43)(0.49,0.53)(0.425,0.2)(0.2,0.47)(0.22,0.37)(0.12,0.4)(0.35,0.14)(0.45,0.1)}
\put(9.035,3.55){\drawline(0.45,0.1)(0.52,0.17)(0.59,0.5)(0.52,0.43)(0.49,0.53)(0.425,0.2)(0.2,0.47)(0.22,0.37)(0.12,0.4)(0.35,0.14)(0.45,0.1)}
\put(9.08,3.4){\drawline(0.45,0.1)(0.52,0.17)(0.59,0.5)(0.52,0.43)(0.49,0.53)(0.425,0.2)(0.2,0.47)(0.22,0.37)(0.12,0.4)(0.35,0.14)(0.45,0.1)}

\put(9.66,5.2){\drawline(0.355,0.1)(0.45,0.136)(0.68,0.4)(0.59,0.37)(0.61,0.47)(0.385,0.21)(0.32,0.53)(0.29,0.43)(0.225,0.5)(0.29,0.148)(0.355,0.1)}
\put(9.615,5.05){\drawline(0.355,0.1)(0.45,0.136)(0.68,0.4)(0.59,0.37)(0.61,0.47)(0.385,0.21)(0.32,0.53)(0.29,0.43)(0.225,0.5)(0.29,0.148)(0.355,0.1)}
\put(9.57,4.9){\drawline(0.355,0.1)(0.45,0.136)(0.68,0.4)(0.59,0.37)(0.61,0.47)(0.385,0.21)(0.32,0.53)(0.29,0.43)(0.225,0.5)(0.29,0.148)(0.355,0.1)}
\put(9.525,4.75){\drawline(0.355,0.1)(0.45,0.136)(0.68,0.4)(0.59,0.37)(0.61,0.47)(0.385,0.21)(0.32,0.53)(0.29,0.43)(0.225,0.5)(0.29,0.148)(0.355,0.1)}
\put(9.48,4.6){\drawline(0.355,0.1)(0.45,0.136)(0.68,0.4)(0.59,0.37)(0.61,0.47)(0.385,0.21)(0.32,0.53)(0.29,0.43)(0.225,0.5)(0.29,0.148)(0.355,0.1)}
\put(9.435,4.45){\drawline(0.355,0.1)(0.45,0.136)(0.68,0.4)(0.59,0.37)(0.61,0.47)(0.385,0.21)(0.32,0.53)(0.29,0.43)(0.225,0.5)(0.29,0.148)(0.355,0.1)}
\put(9.39,4.3){\drawline(0.355,0.1)(0.45,0.136)(0.68,0.4)(0.59,0.37)(0.61,0.47)(0.385,0.21)(0.32,0.53)(0.29,0.43)(0.225,0.5)(0.29,0.148)(0.355,0.1)}
\put(9.345,4.15){\drawline(0.355,0.1)(0.45,0.136)(0.68,0.4)(0.59,0.37)(0.61,0.47)(0.385,0.21)(0.32,0.53)(0.29,0.43)(0.225,0.5)(0.29,0.148)(0.355,0.1)}
\put(9.3,4){\drawline(0.355,0.1)(0.45,0.136)(0.68,0.4)(0.59,0.37)(0.61,0.47)(0.385,0.21)(0.32,0.53)(0.29,0.43)(0.225,0.5)(0.29,0.148)(0.355,0.1)}
\put(9.255,3.85){\drawline(0.355,0.1)(0.45,0.136)(0.68,0.4)(0.59,0.37)(0.61,0.47)(0.385,0.21)(0.32,0.53)(0.29,0.43)(0.225,0.5)(0.29,0.148)(0.355,0.1)}
\put(9.21,3.7){\drawline(0.355,0.1)(0.45,0.136)(0.68,0.4)(0.59,0.37)(0.61,0.47)(0.385,0.21)(0.32,0.53)(0.29,0.43)(0.225,0.5)(0.29,0.148)(0.355,0.1)}
\put(9.165,3.55){\drawline(0.355,0.1)(0.45,0.136)(0.68,0.4)(0.59,0.37)(0.61,0.47)(0.385,0.21)(0.32,0.53)(0.29,0.43)(0.225,0.5)(0.29,0.148)(0.355,0.1)}
\put(9.12,3.4){\drawline(0.355,.1)(.45,.136)(.68,.4)(.59,.37)(.61,.47)(.385,.21)(.32,.53)(.29,.43)(.225,.5)(.29,.148)(.355,.1)}

 \put(8,4){$\Rightarrow$}
\put(4,4){$\Rightarrow$}\put(.8,4){$\Rightarrow$}

\put(-0.5,2.5){\small Point}

\put(2,2.5){\small Step 1}

\put(5.5,2.5){\small Step 2}

\put(9,2.5){\small Step 3}

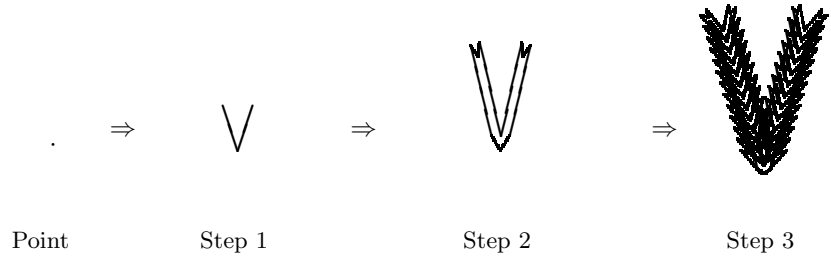
 \put(0,1){\small Fig.10a - One illustration of classical point in fractal manifold after 3 steps}
 \thicklines
\end{picture}

{\bf Remark.} If we use definition \ref{D2}, an element of the
$\varepsilon$-manifold $M_\varepsilon$ corresponds to two points
(with a classical notion of point). If we take into account the
corollary \ref{Cor1}, an element of the $\varepsilon$-manifold
$M_\varepsilon$ becomes a double set of points of the form
(\ref{E2}). The procedure given by corollary \ref{Cor1} is a
transformation of one string of length $L$ to one surface (see
Fig. 10b), then we have appearance of new structure.

 \unitlength=1.2cm
\begin{picture}(6,6)
 \thicklines

 \put(0.5,2.5){\small\line(0,1){.68}}
 \put(2.52,3.7){\small $\sqrt{3}$}


 \put(6.8,5.09){$\bigvee$}
 \put(6.8,4.89){$\bigvee$}
 \put(6.8,4.69){$\bigvee$}
 \put(6.8,4.49){$\bigvee$}
 \put(6.8,4.29){$\bigvee$}
 \put(6.8,4.09){$\bigvee$}
 \put(6.8,3.89){$\bigvee$}

\put(5,1.8){\small Natural construction}

\put(0.3,2){\small Using double mean of $f$ }

\put(0.3,1.7){\small$\forall
(\varepsilon,\delta)\in]0,1]\times[0,1]$}

 \put(6.8,3.69){$\bigvee$}
 \put(6.8,3.49){$\bigvee$}
 \put(6.8,3.29){$\bigvee$}
 \put(6.8,3.09){$\bigvee$}
 \put(6.8,2.89){$\bigvee$}
 \put(6.8,2.69){$\bigvee$}
 \put(6.8,2.49){$\bigvee$}


\put(2.55,3.14){$\bigvee$} \put(5.2,3.79){$\bigvee$}
\put(2.55,3.09){$\bigvee$} \put(5.2,3.69){$\bigvee$}
\put(2.55,3.04){$\bigvee$} \put(5.2,3.59){$\bigvee$}
\put(2.55,2.99){$\bigvee$} \put(5.2,3.49){$\bigvee$}
\put(2.55,2.94){$\bigvee$} \put(5.2,3.39){$\bigvee$}
\put(2.55,2.89){$\bigvee$} \put(5.2,3.29){$\bigvee$}
\put(2.55,2.84){$\bigvee$} \put(5.2,3.19){$\bigvee$}

\put(2.55,2.79){$\bigvee$} \put(5.2,3.09){$\bigvee$}
\put(2.55,2.74){$\bigvee$} \put(5.2,2.99){$\bigvee$}
\put(2.55,2.69){$\bigvee$} \put(5.2,2.89){$\bigvee$}
\put(2.55,2.64){$\bigvee$} \put(5.2,2.79){$\bigvee$}
\put(2.55,2.59){$\bigvee$} \put(5.2,2.69){$\bigvee$}
\put(2.55,2.54){$\bigvee$} \put(5.2,2.59){$\bigvee$}
\put(2.55,2.49){$\bigvee$} \put(5.2,2.49){$\bigvee$}

\put(1.32,2.59){$\Longrightarrow$}

\put(4.32,2.59){$\Longleftarrow$}

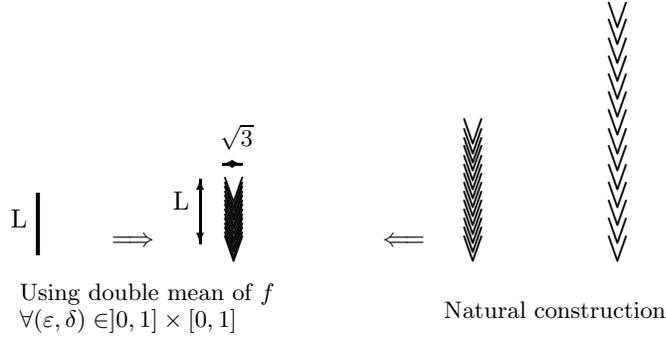
 \put(0,0.5){\small Figure 10b. One illustration of one string in fractal manifold M as represented}

\put(1.5,0.2){\small{in $\bigcup _{\varepsilon \in {\cal
R}_f}\bigcup _{\delta \in {\cal
R}_g}\prod_{i=1}^{3}\Gamma_{i{\delta}}^{\sigma}\times
\{\delta\}\times\{\varepsilon\} $}.}
 \thicklines
 \thinlines
 \put(2.3,2.66){\vector(0,1){.68}}
 \put(2.3,3.3){\vector(0,-1){.68}}
 \put(2,3){L}
 \put(0.2,2.8){L}
 \put(2.56,3.5){\vector(1,0){.2}}
 \put(2.75,3.5){\vector(-1,0){.2}}
\end{picture}
\pesp

\subsection{Substructure}
The fractal manifold deserve this name because of the appearance
of new structure at every step. Indeed

\subsubsection{Step 1: Mean of nowhere differentiable functions}

For the nowhere differentiable function $f_i$, i=1,2,3, introduced
before, the definition \ref{D2} gives us the following diagram
where $\varphi_1=(\varphi_{\varepsilon})_\varepsilon$,
$T_1=(T_{\varepsilon})_\varepsilon$, and
$T_1\circ\varphi_1=(T_\varepsilon\circ\varphi_\varepsilon)_\varepsilon$

\unitlength=0.7cm
\begin{picture}(6,6)

\put(0,4.9){$M=\bigcup _{\varepsilon \in {\cal
Q}_f}M_\varepsilon$}


\put(4.6,4.45){\vector(3,-1){4.7}}

\put(4.6,4.8){\vector(2,0){5.}}

\put(10,4.5){\vector(0,-1){1.5}}


\put(7,5.2){$\varphi_{1}$}

\put(10.5,3.6){$T_1$}

\put(4,3.3){$T_1\circ \varphi _1$}


\put(10,4.8){$\bigcup _{\varepsilon \in {\cal
R}_f}\prod_{i=1}^{3}\Gamma_{i\varepsilon}^{+}\times
\{\varepsilon\} $}

\put(10,2.2){$\bigcup _{\varepsilon \in {\cal
R}_f}\prod_{i=1}^{3}\Gamma_{i\varepsilon}^{-}\times
\{\varepsilon\} $,}

 \thicklines
\end{picture}

We can obtain the same diagram if we use an unknown differentiable
function. The diagram denoted before for the fractal manifold
represents actually a family of diagram of $\varepsilon$-manifold
for all ${\varepsilon\in{\cal R}_f}$.

\subsubsection{Step 2: Mean of the mean}

 Using the process mean of
the mean to approximate the mean functions $f_i(x,\varepsilon)$ of
$f_i$ by another family of mean functions $g_i(x,\delta_1)$ (we
just replace $\prod_{i=1}^3\Gamma_{i0}$ by
$\prod_{i=1}^3\Gamma^{\sigma_1}_{i\varepsilon}$ in corollary
\ref{Cor1}) for $\sigma_1=\pm$, we obtain a second diagram:

\unitlength=0.6cm
\begin{picture}(6,6)

\put(0,4.9){$M_\varepsilon=\prod_{i=1}^3\Gamma^{\sigma_1}_{i\varepsilon}\times\{\varepsilon\}$}


\put(5.1,4.45){\vector(2,-1){3.7}}

\put(5,4.8){\vector(2,0){3.}}

\put(9.3,4.5){\vector(0,-1){1.5}}


\put(5.7,5.2){$(\varphi_{\delta_1})^{\sigma_1}_{{\delta_1},\varepsilon}$}

\put(9.5,3.6){$(T_{\delta_1})^{\sigma_1}_{\delta_1,\varepsilon}$}

\put(4,3.){$(T_{\delta_1} \circ
\varphi_{\delta_1})^{\sigma_1}_{{\delta_1},\varepsilon}$}


\put(8.3,4.8){$\bigcup _{{\delta_1} \in {\cal
R}_{\delta_{1}}}\prod_{i=1}^{3}\Gamma_{i{\delta_1}}^{\sigma_1,+}\times
\{{\delta_1}\}\times\{\varepsilon\} $}

\put(8.3,2){$\bigcup _{{\delta_1} \in {\cal
R}_{\delta_1}}\prod_{i=1}^{3}\Gamma_{i{\delta_1}}^{\sigma_1,-}\times
\{{\delta_1}\}\times\{\varepsilon\} ,$}

 \thicklines
\end{picture}

\ni where for $\sigma_1=\pm$,
$(\varphi_{\delta_1})^{\sigma_1}_{{\delta_1},\varepsilon}$
represents two families
$(\varphi_{\delta_1})^{\sigma_1}_{\delta_1\in{\cal R}_{\delta_1}}$
 at the resolution $\varepsilon$ (respectively for $(T_{\delta_1})^{\sigma_1}_{\delta_1,\varepsilon}$
and $(T_{\delta_1} \circ
\varphi_{\delta_1})^{\sigma_1}_{{\delta_1},\varepsilon}$), with
${\cal R}_{\delta_1}=[0,1]$. The union over $\varepsilon\in{\cal
R}_f$ gives us a new diagram :

\unitlength=0.6cm
\begin{picture}(6,6)

\put(-.5,4.9){$\bigcup_{{\varepsilon} \in {\cal
R}_f}\prod_{i=1}^3\Gamma^{\sigma_1}_{i\varepsilon}\times\{\varepsilon\}$}


\put(5.2,4.45){\vector(2,-1){3.7}}

\put(5.7,4.8){\vector(2,0){3.}}

\put(9.3,4.5){\vector(0,-1){1.5}}


\put(6.5,5.2){$\varphi_2^{\sigma_1}$}

\put(10.5,3.6){$T_2^{\sigma_1}$}

\put(3.7,3.){$(T_2 \circ \varphi _2)^{\sigma_1}$}


\put(8.8,4.8){$\bigcup _{{\varepsilon} \in {\cal R}_f}\bigcup
_{{\delta_1} \in {\cal
R}_{\delta_1}}\prod_{i=1}^{3}\Gamma_{i{\delta_1}}^{\sigma_1,+}\times
\{{\delta_1}\}\times\{\varepsilon\} $}

\put(8.8,2){$\bigcup _{{\varepsilon} \in {\cal R}_f}\bigcup
_{{\delta_1} \in {\cal
R}_{\delta_1}}\prod_{i=1}^{3}\Gamma_{i{\delta_1}}^{\sigma_1,-}\times
\{{\delta_1}\}\times\{\varepsilon\} ,$}

 \thicklines
\end{picture}

\ni where
$\varphi^{\sigma_1}_2=((\varphi_{\delta_1})^{\sigma_1}_{{\delta_1},\varepsilon})_\varepsilon$,
$(T_2\circ\varphi_2)^{\sigma_1}=((T_{\delta_1}\circ\varphi_{\delta_1})^{\sigma_1}_{\delta_1,
\varepsilon})_\varepsilon$, and
$T^{\sigma_1}_2=((T_{\delta_1})^{\sigma_1}_{{\delta_1},\varepsilon})_\varepsilon$,
$\sigma_1=\pm$. Using the diagram of the step 1, we obtain two new
diagrams for the fractal manifold M (at this step we observe an
appearance of a new structure \cite{HU,MA2,TR}):

\unitlength=0.7cm
\begin{picture}(6,6)

\put(0,4.9){$M$}


\put(1.5,4.3){\vector(3,-1){4.5}}

\put(1.5,4.8){\vector(2,0){4.}}

\put(7.5,4.3){\vector(0,-1){1.5}}


\put(3,5.2){$\varphi_2^{+}\circ\varphi_1$}

\put(8.5,3.6){$T_2^{+}$}

\put(.7,2.8){$(T_2\circ\varphi_2)^{+}\circ\varphi_1$}


\put(6,4.8){$\bigcup _{{\varepsilon} \in {\cal R}_f}\bigcup
_{{\delta_1} \in {\cal
R}_{\delta_1}}\prod_{i=1}^{3}\Gamma_{i{\delta_1}}^{+,+}\times
\{{\delta_1}\}\times\{\varepsilon\} $}

\put(6,2){$\bigcup _{{\varepsilon} \in {\cal R}_f}\bigcup
_{{\delta_1} \in {\cal
R}_{\delta_1}}\prod_{i=1}^{3}\Gamma_{i{\delta_1}}^{+,-}\times
\{{\delta_1}\}\times\{\varepsilon\} $,} \thicklines
\end{picture}

\ni and we have  \esp\gesp
 \unitlength=0.7cm
\begin{picture}(5,5)
\put(0,4.5){$M$}


\put(1,4.3){\vector(3,-1){4.7}}

\put(1,4.6){\vector(2,0){4.}}

\put(7.3,4.3){\vector(0,-1){1.5}}


\put(2,5.2){$\varphi_2^{-}\circ T_1\circ\varphi_1$}

\put(7.5,3.6){$T_2^{-}$}

\put(1,2.2){$(T_2\circ\varphi_2)^{-}\circ T_1\circ\varphi_1$}


\put(6,4.8){$\bigcup _{{\varepsilon} \in {\cal R}_f}\bigcup
_{{\delta_1} \in {\cal
R}_{\delta_1}}\prod_{i=1}^{3}\Gamma_{i{\delta_1}}^{-,+}\times
\{{\delta_1}\}\times\{\varepsilon\} $}

\put(6,2){$\bigcup _{{\varepsilon} \in {\cal R}_f}\bigcup
_{{\delta_1} \in {\cal
R}_{\delta_1}}\prod_{i=1}^{3}\Gamma_{i{\delta_1}}^{-,-}\times
\{{\delta_1}\}\times\{\varepsilon\} $.} \thicklines
\end{picture}

\subsubsection{Step n: Mean of the mean n times }

By induction over $n\geq 1$ and using the step 2, we will find
$2^{(n-1)}$ diagrams of the form given by Fig.11, where
$\varphi_k$ is a family of homeomorphisms for $2^{n-1}\leq k\leq
2^{n}-1$, $T_k$ is a family of translations for $2^{n-1}\leq k\leq
2^{n}-1$, and where $\sigma_j=\pm$ for $j=1,..,n-2$. We know that
$\varphi_k$, $T_k$ are families which depend on
$\varepsilon,\delta_1,....,\delta_{n-1}$ for all $2^{n-1}\leq
k\leq 2^{n}-1$, we denoted them with only one index k just for
simplicity. The number n represents the steps of magnification,
and k represents the number of diagrams at a given step.

\unitlength=0.7cm
\begin{picture}(6,6)

\put(-.5,4.9){$M$}


\put(.1,4.5){\vector(1,-1){2}}

\put(0,4.8){\vector(2,0){1.4}}

\put(8.3,4){\vector(0,-1){1}}


\put(.6,5){$\varphi_{k}$}

\put(7.5,3.4){$T_k$}

\put(-.5,3){$T_k \circ\varphi_k$}


\put(1.5,4.8){$\di\bigcup _{{\varepsilon} \in {\cal R}_f}
\di\bigcup_{{\delta_1}\in {\cal R}_{\delta_1}}..
\di\bigcup_{{\delta_{n-1}}\in {\cal R}_{\delta_{n-1}}}
\prod_{i=1}^{3}\Gamma_{i{\delta_{n-1}}}^{{\sigma_1}...{\sigma_{n-2}+}}\times\{\delta_{n-1}\}
\times\dots\times\{\delta_1\}\times\{\varepsilon\} $}

\put(1.5,2){$\di\bigcup _{{\varepsilon} \in {\cal R}_f}
\di\bigcup_{{\delta_1}\in {\cal R}_{\delta_1}}..
\di\bigcup_{{\delta_{n-1}}\in {\cal R}_{\delta_{n-1}}}
\prod_{i=1}^{3}\Gamma_{i{\delta_{n-1}}}^{{\sigma_1}...{\sigma_{n-2}-}}\times\{\delta_{n-1}\}
\times\dots\times\{\delta_1\}\times\{\varepsilon\} $}

\put(7,0){Figure 11.}
 \thicklines
\end{picture}
\gesp

For all object $P$ of $M$ we have $ \varphi_k(P)=\Big(x_{1k},
y_{1k},x_{2k},y_{2k}, x_{3k},y_{3k}\Big), $ for $2^{n-1}\leq k\leq
2^{n}-1$, where in the local chart
$x_{ik}=x_{ik}(\varepsilon,\delta_1,..,\delta_{n-1})$ and
$y_{ik}=y_{ik}(\varepsilon,\delta_1,..,\delta_{n-1})$ given by
\begin{equation}\label{Cor3}
y_{ik}={1\over\delta_{n-1}..\delta_1\varepsilon}
\int_{x_{ik}(\varepsilon,\delta_1,..,\delta_{n-1})}^{x_{ik}(\varepsilon,\delta_1,..,\delta_{n-1})+\delta_{n-1}}
\int_{t_{n-1}}^{t_{n-1}+\delta_{n-2}}..\int_{t_1}^{t_1+\varepsilon}f(t_0)
d t_{0}..d t_{n-1},
\end{equation}
for i=1,2,3, $\forall\delta_1,..,\delta_{n-1}\in[0,1]$,
$\forall\varepsilon\in{\cal R}_{f}$, and $\sigma_j=+$ for example.
For the value $\sigma_j=-$, one have to consider
$\int^{t_j}_{t_j-\delta_{j+1}}$ instead of
$\int_{t_1}^{t_1+\delta_{j+1}}$ in the formula (\ref{Cor3}). The
translation $T_k$ is given by
\begin{equation}\label{TR}
T_k(x_{1k},y_{1k}, x_{2k},y_{2k},
x_{3k},y_{3k})=(x_{1k}+\delta_{n-1},y_{1k},
x_{2k}+\delta_{n-1},y_{2k}, x_{3k}+\delta_{n-1},y_{3k}),
\end{equation}
 and we have the following theorem

\begin{theorem}\label{Th1}
If $M$ is a fractal manifold introduced by definition \ref{D2},
then $\forall n>1$, there exist a family of homeomorphisms
$\varphi_k$, and a family of translations $T_k$ for $2^{n-1}\leq
k\leq 2^{n}-1$, such that one has the $2^{n-1}$ diagrams given
Fig.11.
\end{theorem}
\begin{proof} The result is obtained by induction over the steps n.
\end{proof}

{\bf Remark.} 1) The step 2 corresponds to the procedure given by
corollary \ref{Cor1} which is a transformation of one string of
length $L$ to one surface. To make a smaller transformation,
 we can use the lemma \ref{L2} and then we will obtain a transformation of one string of
length $L$ to one smaller surface.

2) The transformation does not affect the length $L$ of the string
for a given small resolution domain ${\cal R}_f$ of a nowhere
differentiable function. If ${\cal R}_f=]0,1]$ the we will have
$L=1$. We can choose the length of the string as smaller as we
want by ${\cal R}_f=]0,\alpha]$, where $\alpha$ is a small real
number $\alpha\ll 1$ and then we obtain $L=\alpha$.

3) In general a fractal manifold provide a variable tree (See
Fig.12), and using Theorem \ref{Th1} we can see that the charts of
the fractal manifold are changing at every step. Following the
given tree we can see that

In the step 1, the local chart is given by a triplet
$(\Omega,\varphi_1,T_1\circ\varphi_1)$.

In the step 2, the local chart is given by the quintuplet
$(\Omega,\psi_1,\psi_2,\psi_3,\psi_4)$ given by
$(\Omega,\varphi_2\circ\varphi_1,T_2\circ\varphi_2\circ
\varphi_1,\varphi_3\circ T_1\circ\varphi_1,T_3\circ\varphi_3\circ
T_1\circ\varphi_1)$.. etc. That is why we call it a fractal
manifold.

In the diagram given by Fig.12 the following notation are used for
simplicity

i) $ N^{\sigma_1}_{\delta_0}=
\prod_{i=1}^{3}\Gamma_{i{\delta_{0}}}^{\sigma_1}\times\{\delta_0\},
\quad \sigma_1=\pm $, and the integrals used in this sets are
given by $y^+_i=\di{1\over\delta_{0}}
\int_{x_i(\delta_0)}^{x_i(\delta_0)+\delta_{0}} f(t) d t,$ and
$y^-_i=\di{1\over\delta_{0}}
\int^{x_i(\delta_0)}_{x_i(\delta_0)-\delta_{0}} f(t) d t.$

ii)
$N^{\sigma_1\sigma_2}_{\delta_0\delta_1}=\prod_{i=1}^{3}\Gamma_{i{\delta_{1}}}^{\sigma_1\sigma_2}\times
\{{\delta_{1}}\}\times\{\delta_0\}, \quad
(\sigma_1,\sigma_2)=(\pm,\pm)$, and the integrals used in this
sets are given by $$y_{i1}={1\over\delta_{0}\delta_1}
\int_{x_{i1}(\delta_0,\delta_{1})}^{x_{i1}(\delta_0,\delta_{1})+\delta_{1}}
\int_{t_{1}}^{t_{1}+\delta_{0}}f(t_0) d t_{0}d
t_{1},\qquad\hbox{for}\quad (\sigma_1,\sigma_2)=(+,+),$$

$$y_{i2}={1\over\delta_{0}\delta_1}
\int_{x_{i2}(\delta_0,\delta_{1})}^{x_{i2}(\delta_0,\delta_{1})+\delta_{1}}
\int^{t_{1}}_{t_{1}-\delta_{0}}f(t_0) d t_{0}d
t_{1},\qquad\hbox{for}\quad (\sigma_1,\sigma_2)=(+,-),$$

$$y_{i3}={1\over\delta_{0}\delta_1}
\int^{x_{i3}(\delta_0,\delta_{1})}_{x_{i3}(\delta_0,\delta_{1})-\delta_{1}}
\int_{t_{1}}^{t_{1}+\delta_{0}}f(t_0) d t_{0}d
t_{1},\qquad\hbox{for}\quad (\sigma_1,\sigma_2)=(-,+),$$

$$y_{i4}={1\over\delta_{0}\delta_1}
\int^{x_{i4}(\delta_0,\delta_{1})}_{x_{i4}(\delta_0,\delta_{1})-\delta_{1}}
\int^{t_{1}}_{t_{1}-\delta_{0}}f(t_0) d t_{0}d
t_{1},\qquad\hbox{for}\quad (\sigma_1,\sigma_2)=(-,-),$$ etc...

\vskip1.5cm \unitlength=1.2cm
\begin{picture}(7,7)

\put(3,7.2){$\fbox {\bf Fractal Manifold}$}
\put(2.5,5.7){$\bigcup_{\delta_0} N^+_{\delta_0}$}

\put(5,5.7){$\bigcup_{\delta_0} N^-_{\delta_0}$}


\put(0.5,3.5){$\di\bigcup_{\delta_1\delta_0}
N^{++}_{\delta_0\delta_1}$}

\put(2.5,3.5){$\di\bigcup_{\delta_1\delta_0}
N^{+-}_{\delta_0\delta_1}$}

\put(4.9,3.5){$\di\bigcup_{\delta_1\delta_0}
N^{-+}_{\delta_0\delta_1}$}

\put(6.9,3.5){$\di\bigcup_{\delta_1\delta_0}
N^{--}_{\delta_0\delta_1}$}


\put(-0.4,1){$\di\bigcup_{\delta_2\delta_1\delta_0}
N^{+++}_{\delta_0\delta_1\delta_2}$}

\put(1,-0.3){${\di\bigcup_{\delta_2\delta_1\delta_0}
N^{++-}_{\delta_0\delta_1\delta_2}}$}

\put(2,1){$\di\bigcup_{\delta_2\delta_1\delta_0}
N^{+-+}_{\delta_0\delta_1\delta_2}$}

\put(3,-0.3){${\di\bigcup_{\delta_2\delta_1\delta_0}
N^{+--}_{\delta_0\delta_1\delta_2}}$}

\put(4,1){$\di\bigcup_{\delta_2\delta_1\delta_0}
N^{-++}_{\delta_0\delta_1\delta_2}$}

\put(5.3,-0.3){${\di\bigcup_{\delta_2\delta_1\delta_0}
N^{-+-}_{\delta_0\delta_1\delta_2}}$}

\put(6.3,1){$\di\bigcup_{\delta_2\delta_1\delta_0}
N^{--+}_{\delta_0\delta_1\delta_2}$}

\put(7.5,-0.3){${\di\bigcup_{\delta_2\delta_1\delta_0}
N^{---}_{\delta_0\delta_1\delta_2}}$}

\put(9,-1.3){$\vdots$}

\put(8.8,-1.3){$\vdots$}

 \put(8.5,-1.3){$\vdots$}

 \put(8.3,-1.3){$\vdots$}

 \put(7.2,-1.3){$\vdots$}

 \put(7,-1.3){$\vdots$}

 \put(6.7,-1.3){$\vdots$}

 \put(6.5,-1.3){$\vdots$}

 \put(0.2,-1.3){$\vdots$}

 \put(0,-1.3){$\vdots$}

 \put(0.7,-1.3){$\vdots$}

 \put(0.5,-1.3){$\vdots$}

 \put(1.7,-1.3){$\vdots$}

 \put(1.5,-1.3){$\vdots$}

\put(2.2,-1.3){$\vdots$}

 \put(2.,-1.3){$\vdots$}

\put(3,-1.3){$\ldots\ldots\ldots\ldots\ldots\ldots\ldots\ldots$}

\put(4.2,7){\vector(-1,-1){1}}

\put(4.4,7){\vector(1,-1){1}}


\put(2.8,5.5){\vector(-1,-1){1.6}}

\put(3.1,5.5){\vector(0,-3){1.5}}

\put(5.4,5.5){\vector(0,-3){1.5}}

\put(5.8,5.5){\vector(1,-1){1.6}}


\put(1.1,3.2){\vector(-1,-3){.6}}

\put(1.1,3.2){\vector(1,-3){1}}


\put(3.1,3.2){\vector(-1,-3){.6}}

\put(3.1,3.2){\vector(1,-3){1}}


\put(5.4,3.2){\vector(-1,-3){.6}}

\put(5.4,3.2){\vector(1,-3){1}}


\put(7.5,3.2){\vector(-1,-3){.6}}

\put(7.5,3.2){\vector(1,-3){1}}
\put(3.5,5.7){\vector(1,0){1.4}}

\put(1.5,4){\vector(1,0){1.5}}

\put(5.6,4){\vector(1,0){1.5}}

\put(.4,.5){\vector(1,0){.9}}

\put(2.5,.5){\vector(1,0){.9}}

\put(4.7,.5){\vector(1,0){.9}}

\put(7,.5){\vector(1,0){.9}}


 \put(3.2,6.5){$\varphi_1$}

 \put(5.2,6.5){$T_1\circ\varphi_1$}
 \put(1.5,4.8){$\varphi_2$}

 \put(2.8,4.8){$T_2\circ\varphi_2$}

 \put(5,4.8){$\varphi_3$}

 \put(6.9,4.8){$T_3\circ\varphi_3$}

\put(.5,2.5){$\varphi_4$}

\put(1.5,2.1){$T_4\circ\varphi_4$}

\put(2.5,2.5){$\varphi_5$}

\put(3.5,2.1){$T_5\circ\varphi_5$}

\put(4.7,2.5){$\varphi_{6}$}

\put(5.8,2.1){$T_6\circ\varphi_{6}$}

\put(6.8,2.5){$\varphi_{7}$}

\put(7.9,2.1){$T_7\circ\varphi_{7}$}

\put(4.2,5.8){$T_{1}$}

\put(2.2,4.2){$T_{2}$}

\put(5.9,4.2){$T_{3}$}

\put(.8,.2){$T_{4}$}

\put(2.8,.2){$T_{5}$}

\put(4.9,.2){$T_{6}$}

\put(7.3,.2){$T_{8}$}


\put(1.5,-2.3){Figure 12. A fractal diagram of a fractal manifold}
 \thicklines
\end{picture}
\vskip3.5cm

\subsection{Relationship with classical point}

We start this part by an example of fractal manifold.

{\bf Example.} Let $\Gamma_{\varepsilon\over2}$ be the graph of
the mean function $f(x,{\varepsilon\over2})$ of a differentiable
function $f$ at $\varepsilon\over2$ resolution, and
$\Gamma_{\varepsilon}^+$ be the graph of the right mean function
$f(x+{\varepsilon\over2},{\varepsilon\over2})$ of $f$ at
$\varepsilon$ resolution. We define

 $\begin {array}{l} \varphi_\varepsilon :
\Gamma_{\varepsilon\over2} \longrightarrow  \varphi_\varepsilon(\Gamma_{\varepsilon\over2})\\
 (x,f(x,{\varepsilon\over2})) \longmapsto
(x+{\varepsilon\over2},f(x+{\varepsilon\over2},{\varepsilon\over2})
)
\end {array}$\quad and \quad
 $\begin {array}{l} T_{-\varepsilon\over2} :
\varphi_\varepsilon(\Gamma_{\varepsilon\over2}) \longrightarrow  \Gamma_{\varepsilon}^+\\
\qquad (x,y) \longmapsto (x-{\varepsilon\over2},y).
\end {array}$

 We see that:

 1) $\varphi_\varepsilon$ and $T_{-\varepsilon\over2}$ are continuous: each coordinate function
 is continuous, as composite function of continuous functions.

 2) $\varphi^{-1}_\varepsilon$, and $T_{-\varepsilon\over2}^{-1}$ exist, with:
$\varphi^{-1}_\varepsilon(x,y(x,\varepsilon))=(x-{\varepsilon\over2},y(x-{\varepsilon\over2},{\varepsilon\over2}))
$

and $T_{-\varepsilon\over2}^{-1}(x,y)=(x+{\varepsilon\over2},y)$

3) $\varphi^{-1}_\varepsilon$ and $T_{-\varepsilon\over2}^{-1}$
are continuous for the same reason as $\varphi_\varepsilon$ and
$T_{-\varepsilon\over2}$. Then
$T_{-\varepsilon\over2}\circ\varphi_\varepsilon$ is an
homeomorphism from $\Gamma_{\varepsilon\over2}$
 to $\Gamma_\varepsilon^+$. It is not difficult to generalize the
 homeomorphism for the product of three graphs. Conclusion, we obtain an
homeomorphism $\phi_\varepsilon :
\prod_{i=1}^3\Gamma_{i\varepsilon\over2}\times\{\varepsilon\}
\longrightarrow
\prod_{i=1}^3\Gamma_{i\varepsilon}^{+}\times\{\varepsilon\}$, and
then we obtain the following diagram given in Fig.13

\unitlength=0.9cm
\begin{picture}(7,7)

\put(0,4.9){$M_\varepsilon=\prod_{i=1}^3\Gamma_{i\varepsilon\over2}^+\times\{\varepsilon\}$}


\put(4.6,4.45){\vector(2,-1){3.7}}

\put(5,4.8){\vector(2,0){3.}}

\put(8.8,4.5){\vector(0,-1){1.5}}


\put(6,5.2){$\phi_\varepsilon$}

\put(10,3.6){$T_{\varepsilon}$}

\put(3.5,3.){$T_{\varepsilon} \circ \phi_\varepsilon$}


\put(8.5,4.8){$\prod_{i=1}^{3}\Gamma_{i{\varepsilon}}^{+}\times
\{\varepsilon\} $}

\put(8.5,2){$\prod_{i=1}^{3}\Gamma_{i{\varepsilon}}^{-}\times
\{\varepsilon\} .$}

\put(3.5,0.8){\small Figure 13. Diagram of
$\varepsilon\over2$-manifold}
 \thicklines
\end{picture}
\pesp

An internal structure can be found on $\di\bigcup _{\varepsilon
\in {\cal
R}_f}\prod_{i=1}^3\Gamma_{i{\varepsilon\over2}}^+\times\{\varepsilon\}$.

Indeed, $\forall\ P\in\di\bigcup _{\varepsilon \in {\cal
R}_f}\prod_{i=1}^3\Gamma_{i{\varepsilon\over2}}^+\times\{\varepsilon\}$,
there exists $\varepsilon'\in{\cal R}_f$ such that

$P=x(\varepsilon')=\Big(x_1,y(x_1,\varepsilon'),x_2,y(x_2,\varepsilon'),x_3,y(x_3,\varepsilon')\Big)$
with $y(x_i,\varepsilon')=f_i(x_i,{\varepsilon'\over2})$ for
$i=1,2,3$, and where a ${\cal C}^1$ parametric path $x$ is given
by
\begin{equation}
\begin{array}{l} x: {\cal R}_f  \longrightarrow
 \cup_{\varepsilon\in {\cal
R}_f}\prod_{i=1}^3\Gamma_{i{\varepsilon\over2}}^+\times\{\varepsilon\} \\
 \varepsilon
\longmapsto
x(\varepsilon)=\Big(x_1,y(x_1,\varepsilon),x_2,y(x_2,\varepsilon),x_3,y(x_3,\varepsilon)\Big)\in
\prod_{i=1}^3\Gamma_{i{\varepsilon\over2}}^+\times\{\varepsilon\},
\end{array}
\end{equation}
and we have $\forall\varepsilon\in {\cal R}_f,$
$Range(x)\cap\prod_{i=1}^3\Gamma_{i{\varepsilon\over2}}^+\times\{\varepsilon\}=\Big\{x(\varepsilon)\Big\},$
 the $x_i$ are constant and the $y(x_i,\varepsilon)$ are of class
${\cal C}^1$ for $i=1,2,3$. Using the definition \ref{D2}, we
obtain then a fractal manifold

\begin{equation}\label{S5}
M=\di\bigcup _{\varepsilon \in {\cal R}_f}M_\varepsilon=\di\bigcup
_{\varepsilon \in {\cal
R}_f}\prod_{i=1}^3\Gamma_{i\varepsilon\over2}\times\{\varepsilon\},
\end{equation}
and we have the following diagram:

\unitlength=1cm
\begin{picture}(7,7)

\put(0,5.9){$M=\di\bigcup _{\varepsilon \in {\cal
R}_f}\prod_{i=1}^3\Gamma_{i\varepsilon\over2}\times\{\varepsilon\}$}


\put(4,5.45){\vector(2,-1){3}}

\put(4,5.8){\vector(2,0){3.}}

\put(8.3,5.5){\vector(0,-1){1.5}}


\put(4.5,6.2){$(\phi_\varepsilon )_\varepsilon$}

\put(8.5,4.6){$(T_{\varepsilon})_\varepsilon$}

\put(4.5,4.){$(T_{\varepsilon} \circ \phi_\varepsilon
)_\varepsilon$}


\put(7.3,5.8){$\bigcup _{\varepsilon \in {\cal
R}_f}(\prod_{i=1}^{3}\Gamma_{i{\varepsilon}}^{+})\times
\{\varepsilon\} $}

\put(7.3,3.5){$\bigcup _{\varepsilon \in {\cal
R}_f}(\prod_{i=1}^{3}\Gamma_{i{\varepsilon}}^-)\times
\{\varepsilon\}. $}

\put(1.5,0.8){\small Figure 14. Example of fractal manifold
$\bigcup_{\varepsilon\in{\cal R}_f}
\prod_{i=1}^3\Gamma_{i\varepsilon\over2}\times\{\varepsilon\}$}

 \thicklines
\end{picture}
\vskip 2cm

 \unitlength=1.1cm
\begin{picture}(6,6)
 \thicklines
\put(4,4.78){\drawline(1,0.4)(1.12,.6)(1.4,1.8)(1.3,1.66)(1.28,1.84)(1,.6)(1,0.4)}

\put(3.8,4.78){\drawline(1,0.4)(1,.6)(.72,1.84)(.7,1.66)(.6,1.8)(.88,.6)(1,0.4)}
\put(1.5,4.78){\drawline(1.2,1)(1,.4)}
\put(1.4,4.78){\drawline(1,.4)(.8,1)}

\put(-0.3,4.6){\small unreachable}\put(-0.3,4.3){\small point}

\put(4,1.2){\drawline(1,0.4)(1.12,.6)(1.4,1.8)(1.3,1.66)(1.28,1.84)(1,.6)(.72,1.84)(.7,1.66)(.6,1.8)(.88,.6)(1,.4)}
\put(1.5,1.4){\drawline(1.2,1)(1,.4)(.8,1)} \put(0.5,1.8){$\cdot$}

\put(7.14,6.7){\drawline(0.45,0.1)(0.52,0.17)(0.59,0.5)(0.52,0.43)(0.49,0.53)(0.425,0.2)(0.2,0.47)(0.22,0.37)(0.12,0.4)(0.35,0.14)(0.45,0.1)}
\put(7.185,6.55){\drawline(0.45,0.1)(0.52,0.17)(0.59,0.5)(0.52,0.43)(0.49,0.53)(0.425,0.2)(0.2,0.47)(0.22,0.37)(0.12,0.4)(0.35,0.14)(0.45,0.1)}
\put(7.23,6.4){\drawline(0.45,0.1)(0.52,0.17)(0.59,0.5)(0.52,0.43)(0.49,0.53)(0.425,0.2)(0.2,0.47)(0.22,0.37)(0.12,0.4)(0.35,0.14)(0.45,0.1)}
\put(7.275,6.25){\drawline(0.45,0.1)(0.52,0.17)(0.59,0.5)(0.52,0.43)(0.49,0.53)(0.425,0.2)(0.2,0.47)(0.22,0.37)(0.12,0.4)(0.35,0.14)(0.45,0.1)}
\put(7.32,6.1){\drawline(0.45,0.1)(0.52,0.17)(0.59,0.5)(0.52,0.43)(0.49,0.53)(0.425,0.2)(0.2,0.47)(0.22,0.37)(0.12,0.4)(0.35,0.14)(0.45,0.1)}
\put(7.365,5.95){\drawline(0.45,0.1)(0.52,0.17)(0.59,0.5)(0.52,0.43)(0.49,0.53)(0.425,0.2)(0.2,0.47)(0.22,0.37)(0.12,0.4)(0.35,0.14)(0.45,0.1)}
\put(7.41,5.8){\drawline(0.45,0.1)(0.52,0.17)(0.59,0.5)(0.52,0.43)(0.49,0.53)(0.425,0.2)(0.2,0.47)(0.22,0.37)(0.12,0.4)(0.35,0.14)(0.45,0.1)}
\put(7.455,5.65){\drawline(0.45,0.1)(0.52,0.17)(0.59,0.5)(0.52,0.43)(0.49,0.53)(0.425,0.2)(0.2,0.47)(0.22,0.37)(0.12,0.4)(0.35,0.14)(0.45,0.1)}
\put(7.5,5.5){\drawline(0.45,0.1)(0.52,0.17)(0.59,0.5)(0.52,0.43)(0.49,0.53)(0.425,0.2)(0.2,0.47)(0.22,0.37)(0.12,0.4)(0.35,0.14)(0.45,0.1)}
\put(7.545,5.35){\drawline(0.45,0.1)(0.52,0.17)(0.59,0.5)(0.52,0.43)(0.49,0.53)(0.425,0.2)(0.2,0.47)(0.22,0.37)(0.12,0.4)(0.35,0.14)(0.45,0.1)}
\put(7.59,5.2){\drawline(0.45,0.1)(0.52,0.17)(0.59,0.5)(0.52,0.43)(0.49,0.53)(0.425,0.2)(0.2,0.47)(0.22,0.37)(0.12,0.4)(0.35,0.14)(0.45,0.1)}
\put(7.635,5.05){\drawline(0.45,0.1)(0.52,0.17)(0.59,0.5)(0.52,0.43)(0.49,0.53)(0.425,0.2)(0.2,0.47)(0.22,0.37)(0.12,0.4)(0.35,0.14)(0.45,0.1)}
\put(7.68,4.9){\drawline(0.45,0.1)(0.52,0.17)(0.59,0.5)(0.52,0.43)(0.49,0.53)(0.425,0.2)(0.2,0.47)(0.22,0.37)(0.12,0.4)(0.35,0.14)(0.45,0.1)}

\put(8.66,6.7){\drawline(0.355,0.1)(0.45,0.136)(0.68,0.4)(0.59,0.37)(0.61,0.47)(0.385,0.21)(0.32,0.53)(0.29,0.43)(0.225,0.5)(0.29,0.148)(0.355,0.1)}
\put(8.615,6.55){\drawline(0.355,0.1)(0.45,0.136)(0.68,0.4)(0.59,0.37)(0.61,0.47)(0.385,0.21)(0.32,0.53)(0.29,0.43)(0.225,0.5)(0.29,0.148)(0.355,0.1)}
\put(8.57,6.4){\drawline(0.355,0.1)(0.45,0.136)(0.68,0.4)(0.59,0.37)(0.61,0.47)(0.385,0.21)(0.32,0.53)(0.29,0.43)(0.225,0.5)(0.29,0.148)(0.355,0.1)}
\put(8.525,6.25){\drawline(0.355,0.1)(0.45,0.136)(0.68,0.4)(0.59,0.37)(0.61,0.47)(0.385,0.21)(0.32,0.53)(0.29,0.43)(0.225,0.5)(0.29,0.148)(0.355,0.1)}
\put(8.48,6.1){\drawline(0.355,0.1)(0.45,0.136)(0.68,0.4)(0.59,0.37)(0.61,0.47)(0.385,0.21)(0.32,0.53)(0.29,0.43)(0.225,0.5)(0.29,0.148)(0.355,0.1)}
\put(8.435,5.95){\drawline(0.355,0.1)(0.45,0.136)(0.68,0.4)(0.59,0.37)(0.61,0.47)(0.385,0.21)(0.32,0.53)(0.29,0.43)(0.225,0.5)(0.29,0.148)(0.355,0.1)}
\put(8.39,5.8){\drawline(0.355,0.1)(0.45,0.136)(0.68,0.4)(0.59,0.37)(0.61,0.47)(0.385,0.21)(0.32,0.53)(0.29,0.43)(0.225,0.5)(0.29,0.148)(0.355,0.1)}
\put(8.345,5.65){\drawline(0.355,0.1)(0.45,0.136)(0.68,0.4)(0.59,0.37)(0.61,0.47)(0.385,0.21)(0.32,0.53)(0.29,0.43)(0.225,0.5)(0.29,0.148)(0.355,0.1)}
\put(8.3,5.5){\drawline(0.355,0.1)(0.45,0.136)(0.68,0.4)(0.59,0.37)(0.61,0.47)(0.385,0.21)(0.32,0.53)(0.29,0.43)(0.225,0.5)(0.29,0.148)(0.355,0.1)}
\put(8.255,5.35){\drawline(0.355,0.1)(0.45,0.136)(0.68,0.4)(0.59,0.37)(0.61,0.47)(0.385,0.21)(0.32,0.53)(0.29,0.43)(0.225,0.5)(0.29,0.148)(0.355,0.1)}
\put(8.21,5.2){\drawline(0.355,0.1)(0.45,0.136)(0.68,0.4)(0.59,0.37)(0.61,0.47)(0.385,0.21)(0.32,0.53)(0.29,0.43)(0.225,0.5)(0.29,0.148)(0.355,0.1)}
\put(8.165,5.05){\drawline(0.355,0.1)(0.45,0.136)(0.68,0.4)(0.59,0.37)(0.61,0.47)(0.385,0.21)(0.32,0.53)(0.29,0.43)(0.225,0.5)(0.29,0.148)(0.355,0.1)}
\put(8.12,4.9){\drawline(0.355,.1)(.45,.136)(.68,.4)(.59,.37)(.61,.47)(.385,.21)(.32,.53)(.29,.43)(.225,.5)(.29,.148)(.355,.1)}


\put(7.54,3.2){\drawline(0.45,0.1)(0.52,0.17)(0.59,0.5)(0.52,0.43)(0.49,0.53)(0.425,0.2)(0.2,0.47)(0.22,0.37)(0.12,0.4)(0.35,0.14)(0.45,0.1)}
\put(7.585,3.05){\drawline(0.45,0.1)(0.52,0.17)(0.59,0.5)(0.52,0.43)(0.49,0.53)(0.425,0.2)(0.2,0.47)(0.22,0.37)(0.12,0.4)(0.35,0.14)(0.45,0.1)}
\put(7.63,2.9){\drawline(0.45,0.1)(0.52,0.17)(0.59,0.5)(0.52,0.43)(0.49,0.53)(0.425,0.2)(0.2,0.47)(0.22,0.37)(0.12,0.4)(0.35,0.14)(0.45,0.1)}
\put(7.675,2.75){\drawline(0.45,0.1)(0.52,0.17)(0.59,0.5)(0.52,0.43)(0.49,0.53)(0.425,0.2)(0.2,0.47)(0.22,0.37)(0.12,0.4)(0.35,0.14)(0.45,0.1)}
\put(7.72,2.6){\drawline(0.45,0.1)(0.52,0.17)(0.59,0.5)(0.52,0.43)(0.49,0.53)(0.425,0.2)(0.2,0.47)(0.22,0.37)(0.12,0.4)(0.35,0.14)(0.45,0.1)}
\put(7.765,2.45){\drawline(0.45,0.1)(0.52,0.17)(0.59,0.5)(0.52,0.43)(0.49,0.53)(0.425,0.2)(0.2,0.47)(0.22,0.37)(0.12,0.4)(0.35,0.14)(0.45,0.1)}
\put(7.81,2.3){\drawline(0.45,0.1)(0.52,0.17)(0.59,0.5)(0.52,0.43)(0.49,0.53)(0.425,0.2)(0.2,0.47)(0.22,0.37)(0.12,0.4)(0.35,0.14)(0.45,0.1)}
\put(7.855,2.15){\drawline(0.45,0.1)(0.52,0.17)(0.59,0.5)(0.52,0.43)(0.49,0.53)(0.425,0.2)(0.2,0.47)(0.22,0.37)(0.12,0.4)(0.35,0.14)(0.45,0.1)}
\put(7.9,2){\drawline(0.45,0.1)(0.52,0.17)(0.59,0.5)(0.52,0.43)(0.49,0.53)(0.425,0.2)(0.2,0.47)(0.22,0.37)(0.12,0.4)(0.35,0.14)(0.45,0.1)}
\put(7.945,1.85){\drawline(0.45,0.1)(0.52,0.17)(0.59,0.5)(0.52,0.43)(0.49,0.53)(0.425,0.2)(0.2,0.47)(0.22,0.37)(0.12,0.4)(0.35,0.14)(0.45,0.1)}
\put(7.99,1.7){\drawline(0.45,0.1)(0.52,0.17)(0.59,0.5)(0.52,0.43)(0.49,0.53)(0.425,0.2)(0.2,0.47)(0.22,0.37)(0.12,0.4)(0.35,0.14)(0.45,0.1)}
\put(8.035,1.55){\drawline(0.45,0.1)(0.52,0.17)(0.59,0.5)(0.52,0.43)(0.49,0.53)(0.425,0.2)(0.2,0.47)(0.22,0.37)(0.12,0.4)(0.35,0.14)(0.45,0.1)}
\put(8.08,1.4){\drawline(0.45,0.1)(0.52,0.17)(0.59,0.5)(0.52,0.43)(0.49,0.53)(0.425,0.2)(0.2,0.47)(0.22,0.37)(0.12,0.4)(0.35,0.14)(0.45,0.1)}

\put(8.66,3.2){\drawline(0.355,0.1)(0.45,0.136)(0.68,0.4)(0.59,0.37)(0.61,0.47)(0.385,0.21)(0.32,0.53)(0.29,0.43)(0.225,0.5)(0.29,0.148)(0.355,0.1)}
\put(8.615,3.05){\drawline(0.355,0.1)(0.45,0.136)(0.68,0.4)(0.59,0.37)(0.61,0.47)(0.385,0.21)(0.32,0.53)(0.29,0.43)(0.225,0.5)(0.29,0.148)(0.355,0.1)}
\put(8.57,2.9){\drawline(0.355,0.1)(0.45,0.136)(0.68,0.4)(0.59,0.37)(0.61,0.47)(0.385,0.21)(0.32,0.53)(0.29,0.43)(0.225,0.5)(0.29,0.148)(0.355,0.1)}
\put(8.525,2.75){\drawline(0.355,0.1)(0.45,0.136)(0.68,0.4)(0.59,0.37)(0.61,0.47)(0.385,0.21)(0.32,0.53)(0.29,0.43)(0.225,0.5)(0.29,0.148)(0.355,0.1)}
\put(8.48,2.6){\drawline(0.355,0.1)(0.45,0.136)(0.68,0.4)(0.59,0.37)(0.61,0.47)(0.385,0.21)(0.32,0.53)(0.29,0.43)(0.225,0.5)(0.29,0.148)(0.355,0.1)}
\put(8.435,2.45){\drawline(0.355,0.1)(0.45,0.136)(0.68,0.4)(0.59,0.37)(0.61,0.47)(0.385,0.21)(0.32,0.53)(0.29,0.43)(0.225,0.5)(0.29,0.148)(0.355,0.1)}
\put(8.39,2.3){\drawline(0.355,0.1)(0.45,0.136)(0.68,0.4)(0.59,0.37)(0.61,0.47)(0.385,0.21)(0.32,0.53)(0.29,0.43)(0.225,0.5)(0.29,0.148)(0.355,0.1)}
\put(8.345,2.15){\drawline(0.355,0.1)(0.45,0.136)(0.68,0.4)(0.59,0.37)(0.61,0.47)(0.385,0.21)(0.32,0.53)(0.29,0.43)(0.225,0.5)(0.29,0.148)(0.355,0.1)}
\put(8.3,2){\drawline(0.355,0.1)(0.45,0.136)(0.68,0.4)(0.59,0.37)(0.61,0.47)(0.385,0.21)(0.32,0.53)(0.29,0.43)(0.225,0.5)(0.29,0.148)(0.355,0.1)}
\put(8.255,1.85){\drawline(0.355,0.1)(0.45,0.136)(0.68,0.4)(0.59,0.37)(0.61,0.47)(0.385,0.21)(0.32,0.53)(0.29,0.43)(0.225,0.5)(0.29,0.148)(0.355,0.1)}
\put(8.21,1.7){\drawline(0.355,0.1)(0.45,0.136)(0.68,0.4)(0.59,0.37)(0.61,0.47)(0.385,0.21)(0.32,0.53)(0.29,0.43)(0.225,0.5)(0.29,0.148)(0.355,0.1)}
\put(8.165,1.55){\drawline(0.355,0.1)(0.45,0.136)(0.68,0.4)(0.59,0.37)(0.61,0.47)(0.385,0.21)(0.32,0.53)(0.29,0.43)(0.225,0.5)(0.29,0.148)(0.355,0.1)}
\put(8.12,1.4){\drawline(0.355,.1)(.45,.136)(.68,.4)(.59,.37)(.61,.47)(.385,.21)(.32,.53)(.29,.43)(.225,.5)(.29,.148)(.355,.1)}

\put(7,4.3){\small Nowhere differentiable }

\put(7.5,1){\small Differentiable }

\put(-0.3,1){\small classical point}

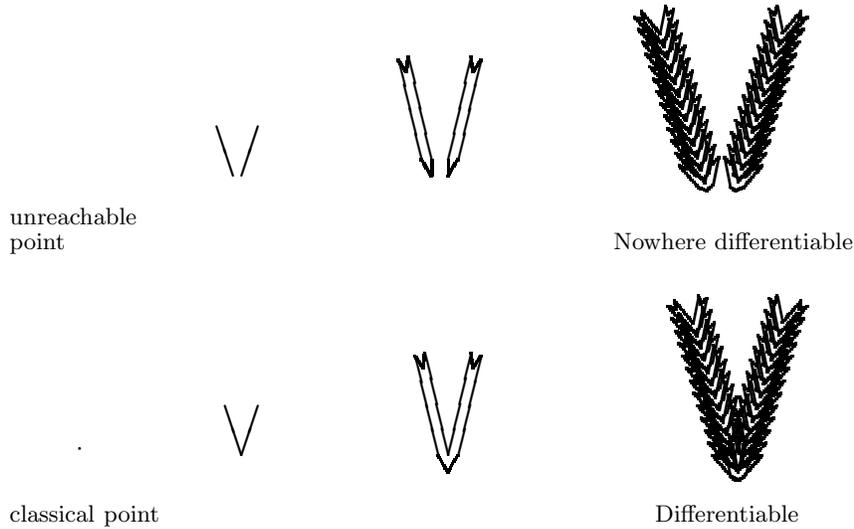
 \put(1,0){\small Figure 15.  One illustration of the process mean of the mean}
 \thicklines
\end{picture}

\gesp
 On one hand, we have
$\prod_{i=1}^3\Gamma_{i\varepsilon\over2}^+\times\{\varepsilon\}$
homeomorphic to $\rR^3$, then the set $M$ given by the formula
(\ref{S5}) is homeomorphic to $\rR^3$, and the elements of this
set are points (classical definition). On another hand, the set
$M$ given by the formula (\ref{S5}) is a fractal manifold, where
the elements are objects given by (\ref{E1}). We observe two kinds
of objects obtained by the process mean of mean: one concerning
nowhere differentiable function, and another one concerning
differentiable function (see Illustration Fig.15). 

As we can see,
the mechanism is the same, the only difference is the appearance
of discontinuity.

If we consider only the mean function (\ref{M}) of $f$ ( and not
the right- left mean function), the same construction leads us to
obtain a differentiable family of manifold given by $M= \cup
_{\varepsilon \in {\cal R}_f} M_{\varepsilon}$, but this family
$(M_\varepsilon)_{\varepsilon\in\rR}$ doesn't give us an
appearance of new structure: indeed, $\forall\varepsilon\in {\cal
R}_f$, we obtain the same local object, and then we have the same
object (only one string) for differentiable and nowhere
differentiable functions.

Following Remark (4.2 Substructure) , the dimension of a fractal
manifold is variable. It varies between 5 and infinity. More
details will be given in \cite{BP} about a non linear analysis on
fractal manifold, and details will be given about the nature of
the expansion into this kind of manifold in \cite{BF1}. An
application in modern cosmology will also be given in \cite{BF2}
using fractal manifold.
We constructed a geometric space which is neither a continuum nor
a discrete space, but a mixture of both. Locally this space is a
disjoint union over $\varepsilon\in]0,\alpha]$ of a family of two
ordinary continuums. Space acquires a variable geometry, namely,
it becomes explicitly dependent on the resolution. However,
following this framework, the geometry of space is assumed to be
characterized not only by curvature, but also by appearance of new
structure at every step. The classical differentiable geometry
doesn't give us useful information of this kind of double space,
however one should construct a non linear analysis over this kind
of manifolds.

{\bf Acknowledgment:} I would like to thank {\it
$H\acute{e}l\grave{e}$ne Porchon} for thoughtful and helpful
discussions, and corrections. Many thanks to {\it Jaouad Sahbani}
for the long hours of discussion, corrections and encouragement.

 \pesp

\begin{small}

\end{small}

\end{document}